\documentstyle[12pt]{article}
\renewcommand{\a}{\alpha}

           \newcommand{\G}{\Gamma}
\renewcommand{\d}{\delta}         \newcommand{\D}{\Delta}
\newcommand{\ve}{\varepsilon}

\newcommand{\k}{\kappa}
\newcommand{\ld}{\lambda}        
\newcommand{\om}{\omega}         \newcommand{\OM}{\Omega}
\newcommand{\p}{\psi}             \newcommand{\PS}{\Psi} 
\newcommand{\ro}{\rho}
           
\newcommand{\th}{\theta}         
\newcommand{\f}{{\phi}}           \newcommand{\F}{{\Phi}}
\newcommand{\vf}{{\varphi}}
       
\newcommand{\z}{\zeta} 
 
\newcommand{\cD}{{\cal D}}

\newcommand{\cF}{{\cal F}}

\newcommand{\cH}{{\cal H}}

\newcommand{\cL}{{\cal L}}
\newcommand{\cM}{{\cal M}}

\newcommand{\cS}{{\cal S}}

\newcommand{\cU}{{\cal U}}

\newcommand{\cX}{{\cal X}}

\newcommand{\hA}{{\widehat A}}

\newcommand{\hU}{{\widehat U}}

\newcommand{\hX}{{\widehat X}}

\newcommand{\Hp}{{\widehat p}}

\newcommand{\Ht}{{\widehat t}}

\newcommand{\Hx}{{\widehat x}}

\newcommand{\bz}{{\bar{z}}}
\newcommand{\bv}{{\bar{v}}}

\newcommand{\bt}{{\bar{t}}}
\newcommand{\bpa}{{\bar{\pa}}}
\newcommand{\bxi}{{\bar{\xi}}}

\newcommand{\bzt}{{\bar{\z}}}
\newcommand{\bN}{\bar{N}}
\newcommand{\deff}{\,\stackrel{\rm def}{\equiv}\,}
\newcommand{\lra}{\longrightarrow}
\newcommand{\ra}{\,\rightarrow\,}
\def\limar#1#2{\,\raise0.3ex\hbox{$\longrightarrow$\kern-1.5em\raise-1.1ex
\hbox{$\scriptstyle{#1\rightarrow #2}$}}\,}
\def\limarr#1#2{\,\raise0.3ex\hbox{$\longrightarrow$\kern-1.5em\raise-1.3ex
\hbox{$\scriptstyle{#1\rightarrow #2}$}}\,}
\def\limlar#1#2{\ \raise0.3ex
\hbox{$-\hspace{-0.5em}-\hspace{-0.5em}-\hspace{-0.5em}
\longrightarrow$\kern-2.7em\raise-1.1ex
\hbox{$\scriptstyle{#1\rightarrow #2}$}}\ \ }

\newcommand{\wt}{\widetilde}

\newcommand{\da}{{\dagger}}

\def\h{\hbar}

\newcommand{\exx}[1]{\exp\left\{ {#1}\right\}}

\newcommand{\one}{{\leavevmode{\rm 1\mkern -5.4mu I}}}
\newcommand{\Z}{Z\!\!\!Z}
\newcommand{\Ibb}[1]{ {\rm I\ifmmode\mkern
            -3.6mu\else\kern -.2em\fi#1}}
\newcommand{\ibb}[1]{\leavevmode\hbox{\kern.3em\vrule
     height 1.2ex depth -.3ex width .2pt\kern-.3em\rm#1}}

\newcommand{\C}{{\ibb C}}
\newcommand{\R}{{\Ibb R}}

\newcommand{\rational}{{\kern .1em {\raise .47ex
\hbox{$\scripscriptstyle |$}}
    \kern -.35em {\rm Q}}}

\newcommand{\Ree}{{\cal R}\!e \,}
\newcommand{\Imm}{{\cal I}\!m \,}
\newcommand{\tr}{{\rm {Tr} \,}}
\newcommand{\er}{{\rm{e}}}
\renewcommand{\i}{{\rm{i}}}

\newcommand{\id}{{\rm{id}\,}}
\newcommand{\ad}{{\rm{ad}\,}}

\newcommand{\const}{{\rm{\,const\,}}}

\newcommand{\pa}{\partial}
\newcommand{\pad}[2]{{\frac{\partial #1}{\partial #2}}}

\newcommand{\eg}{{\em e.g.,\ }}
\newcommand{\ie}{{{\em i.e.}\ }}
\newcommand{\etc}{{\em etc.\ }}

\newcommand{\cf}{{\em cf.\ }}
\def\<{\langle}
\def\>{\rangle}
\def\lgl{\langle\langle}
\def\rgr{\rangle\rangle}
\newcommand{\bra}[1]{\left\langle {#1}\right|}
\newcommand{\ket}[1]{\left| {#1}\right\rangle}

\newcommand{\be}{\begin{equation}}
\newcommand{\ee}{\end{equation}}
\newcommand{\bn}{\begin{eqnarray}}
\newcommand{\en}{\end{eqnarray}}
\newcommand{\bnn}{\begin{eqnarray*}}
\newcommand{\enn}{\end{eqnarray*}}
\newcommand{\ba}{\begin{array}}
\newcommand{\ea}{\end{array}}
\newcommand{\e}{\label}
\newcommand{\nbr}{\nonumber\\[3mm]}
\newcommand{\r}[1]{(\ref{#1})}

\newcommand{\qq}{\qquad}

\newcommand{\biz}{\begin{itemize}} 
\newcommand{\eiz}{\end{itemize}}
\newcommand{\ben}{\begin{enumerate}} 
\newcommand{\een}{\end{enumerate}}

\newcommand{\fr}{Fourier\ }

\def\raczka{a\kern-0.35em\raise-0.4ex\hbox{$\scriptscriptstyle c$}}
\font\goth=eufm10
\def\gg{\hbox{\goth g}}

\def\qint{\int\kern-0.9em\raise2.2ex\hbox{${\textstyle
q}$}\ }
\newcommand{\paq}{\partial_q}
\newcommand{\bpaq}{\bar{\partial}_q}
\begin{document}
\begin{titlepage}
\
\vspace{0.5cm}
\begin{flushright}
{\bf HIP-1998-77/TH}\\
hep-th/9812180
\end{flushright}

\begin{center}
\vspace*{1.0cm}

{\Large \bf Quantum Field Theory on Noncommutative\\ 
\vspace{5mm}
Space-Times and the Persistence of\\
\vspace{5mm}
Ultraviolet Divergences}

\vskip 1cm

{\large {\bf M. Chaichian}}, 
\ \ \ {\large{\bf 
A. Demichev}}\renewcommand{\thefootnote}{a}
\footnote{Permanent address: 
Nuclear Physics Institute, Moscow State University,
119899, Moscow, Russia}\ \ and \ \ 
{\large{\bf 
P. Pre\v{s}najder}}\renewcommand{\thefootnote}{b}
\footnote{Permanent address:
Department of Theoretical Physics, Comenius University,
Mlynsk\'{a} dolina, SK-84215 Bratislava, Slovakia}

\vskip 0.2cm

High Energy Physics Division, Department of Physics,\\
University of Helsinki\\
\ \ {\small\it and}\\
\ \ Helsinki Institute of Physics,\\
P.O. Box 9, FIN-00014 Helsinki, Finland

\end{center}

\vspace{0.5 cm}

\begin{abstract}
\normalsize

We study properties of a scalar quantum field theory on two-dimensional 
noncommutative space-times.  Contrary to the common belief that
noncommutativity of space-time would be a key to remove the ultraviolet
divergences, we show  that field theories on a noncommutative plane with
the most natural Heisenberg-like commutation relations among coordinates  or
even on a noncommutative quantum plane with $E_q(2)$-symmetry have ultraviolet
divergences, while the theory on a noncommutative cylinder is ultraviolet
finite. Thus, ultraviolet behaviour of a field theory on noncommutative spaces
is sensitive to the topology of the space-time, namely to its compactness.  We
present general  arguments for the case of higher space-time dimensions and  as
well discuss the symmetry transformations of physical states on noncommutative
space-times.

\end{abstract}

\vspace{0.5cm}
{\it PACS:}\ \ 03.70

\vspace{0.3cm}

{\it Keywords:}\hspace{0.5cm} 
{\small Quantum field theory, ultraviolet divergences, regularization,\\ 
$\phantom{Keywords:\ \hspace{1cm}}$ noncommutative space-times.}
\end{titlepage}

\section{Introduction}
The standard concept of a geometric space is based on the notion of a manifold
$\cM$ whose points $x\in\cM$ are locally labelled by a finite number of real
coordinates $x^{\mu}\in\R^4$. However, it is generally believed that the
picture of space-time as a manifold $\cM$ should break down at very short 
distances of the order of the
Planck  length $\ld_P\approx 1.6\times 10^{-33}\,cm$. This implies that the
mathematical concepts for high energy (small distance) physics have to be
changed, or more precisely, our classical geometrical concepts may not be well
suited for the description of physical phenomena at small  distances.
No convincing alternative description of physics at very short distances 
is known, though different routes to progress
have been proposed. One such direction is to try to formulate physics on some
noncommutative space-time. There appear to be too many possibilities to do
this, and it is difficult to see what the right choice is. There have been
investigations in the context of Connes' approach \cite{Connes} to gravity 
and the Standard Model of electroweak and strong interactions 
\cite{ConnesL,Chams}, relation between measurements at very small
distances and black hole formations \cite{DoplicherFR} and
string theory \cite{strings-nc}. One more possibility
is based on quantum group theory (see, \eg \cite{ChaichianDbk}).
It is worth to note that the generalization of commutation relations for the 
canonical operators (coordinate-momentum or creation-annihilation operators) 
was suggested long ago by Heisenberg 
\cite{Heisenberg} in attempts to achieve regularization for his 
(nonrenormalizable) nonlinear spinor field theory (see also Wigner 
\cite{Wigner}, and for a review \cite{ChaichianDbk}).

The essence of the noncommutative geometry consists
in reformulating first the geometry in terms of commutative algebras
and modules of smooth functions, and then generalizing them to their
noncommutative analogs. 
If the notions of the noncommutative geometry are used directly for the
description of the space-time, the notion of points as elementary geometrical
entity is lost and one may expect that an ultraviolet (UV) cutoff appears. 
The simplest model of this kind is the fuzzy sphere 
(see \cite{Ber,Hoppe,GKP3} and refs. therein),
\ie the noncommutative analog of a two-dimensional sphere. 
As is well known from the standard quantum mechanics, a quantization of 
any {\it compact} space, in particular a sphere, leads to
finite-dimensional representations of the corresponding operators, so that  
in this case any calculation is reduced to manipulations with 
finite-dimensional matrices and thus there is simply no place for
UV-divergences \cite{GKP3}. It is worth to mention also that the parameter of
noncommutativity (analog of the Planck constant) in the case of quantization 
of compact manifolds is related to dimensionality of a matrix representation
\cite{Ber}. After removing the noncommutative parameter, the known results
of the classical geometry are recovered. 

Things are not so easy in the case of non-compact manifolds. The quantization
leads to infinite-dimensional representations and we have no guarantee that
noncommutativity of the space-time coordinates removes the UV-divergences. A
relatively simple type of noncommutative geometry of non-compact Minkowski 
space with Heisenberg-like commutation relations among coordinates,  has been
considered in \cite{DoplicherFR}. Later it was shown \cite{Filk} that this
model has the same UV-behaviour as an ordinary field theory on the commutative
space-time. In the next section we shall use this example, reduced to
two-dimensional space, for a general discussion of construction of a quantum
field theory on a noncommutative space (NC-QFT) and argue that the underlying
reason for such a UV-behaviour of the model is that the corresponding algebra
of quantum mechanical operators is isomorphic to the usual one and that the
momenta degrees of freedom are associated to the {\it non-compact}
Heisenberg-Weyl  group manifold. In discussing this example, we shall see that
physically meaningful quantities in NC-QFT are the correlation functions (Green
functions)  of {mean values} of
quantum fields on a noncommutative space-time in {\it localized states} from
the Hilbert space of a representation of the corresponding coordinate algebra.
These localized states are the best counterparts of points on an ordinary
commutative space, which "label" (infinite number of ) degrees of freedom of
the QFT. Thus we must consider a quantum field in NC-QFT as a map from the 
{\it set of states}  on the corresponding noncommutative space into the 
algebra of secondary  quantized operators. We also use this example for the
study of symmetry transformations of noncommutative space-times with Lie
algebra commutation relations for coordinates. The noncommutative coordinates
prove to be tensor operators, and we consider concrete examples of the
corresponding transformations of localized states (analog of space-time point
transformations).

In section 3 we consider more complicated example of NC-QFT, defined on a
noncommutative cylinder. The latter is obtained via quantization of its
classical counterpart, by considering it as a co-adjoint orbit of the
two-dimensional  Euclidean group $E(2)$ with the uniquely defined
Lie-Kirillov-Kostant Poisson structure. As is well known, a scalar
two-dimensional QFT on a commutative cylinder has the same UV-properties as QFT
on a plane (in particular, divergent  tadpole Feynman diagram): UV-behaviour is
related to small distances and, hence, insensitive to the topology of a
coordinate manifold. On the  contrary, quantization procedure is very sensitive
to the topology and this leads to essentially different properties of NC-QFT on
the noncommutative cylinder and on the noncommutative plane. While the latter
retains the divergent tadpoles (as an ordinary QFT), the former proves to be
UV-finite. 

It is known that a cylinder is homeomorphic to a plane with one punctured 
point. In the noncommutative case, this map corresponds to transition to a
quantum plane. In section 4 we proceed to the study of a quantum plane with
$E_q(2)$-symmetry. Now the coordinate and momentum operators form a subalgebra
of ordinary quantum mechanical algebra, so that some operators have discrete
spectra. This fact could inspire the belief that the corresponding NC-QFT has 
an improved UV-behaviour. However, the explicit calculation shows that this is
not the case and the QFT on the quantum plane has the usual UV-divergences. 

Section 5 is devoted to the summary of the results. In particular,  the
compactness of a noncommutative space-time is crucial for the UV-behaviour of
NC-QFT, namely,  at most one dimension (time) is allowed to be noncompact, in
order to achieve the removal of UV-divergences of a quantum field theory
formulated in a noncommutative space-time of arbitrary dimensions.

\section{Two-dimensional quantum field theory on noncommutative space-time 
with Heisenberg-like commutation relations}

As an introductory example, we consider the simplest case of a two-dimensional
Euclidean field theory with the noncommutativity of the Heisenberg type. 
However, for the convenience of exposition we start with a brief summary of 
the standard commutative case.

\subsection{Two-dimensional fields on a Euclidean commutative plane}

The complex scalar field $\vf (x)$ on a Euclidean plane $P^{(2)} =\R^2$
is a prescription
$$
x = (x_1 ,x_2 ) \in P^{(2)}\ \to \ \vf (x) \in \C\ ,
$$
which assigns to any point $x$ of the plane the complex number $\vf (x)$.

There is an equivalent description of functions on $P^{(2)}$ in terms of the
Fourier transform:
\be
f(x)\ =\ \frac{1}{(2\pi)^2} \int d^2 k\ {\tilde f}(k) e^{ikx} \ .
\ee
The integral of $f(x)$ over  $P^{(2)}$ can be expressed as
\be
I_0 [f]\ =\ \int d^2 x\ f(x) \ =\  {\tilde f}(0) \ ,
\ee
where $d^2x$ is the standard Lebesgue measure on $P^{(2)}$.
The point-wise multiplication of fields can be expressed
as a convolution of their \fr transforms:
\bnn
f(x) \vf (x)\ &=&\ \int d^2 k\ ({\tilde f}
\circ {\tilde \vf})(k)\er^{\i kx} \ ,\nbr
({\tilde f}\circ {\tilde \vf})(k) \ &\equiv&\ \frac{1}{(2\pi)^2}
\int d^2 q  d^2 q'\ \delta (k-q-q' ) {\tilde f}(q){\tilde \vf}(q')\ .
\enn
The Euclidean action of self-interacting scalar field usually has the form
\be
S[\vf ,\vf^* ]\ =\ S_0 [\vf ,\vf^* ]\ +\ S_{int} [\vf ,\vf^* ]\ ,\e{C-act}
\ee
where
\be
S_0 [\vf ,\vf^* ]=\int d^2 x\ [(\partial_i \vf )^* (\partial_i \vf )
\ +\ m^2 \vf^* \vf]=\frac{1}{(2\pi)^2}\int d^2 k\
{\tilde \vf}^* (k)[k^2 +m^2 ] {\tilde \vf}(k) \ ,   \e{clas.free.act}
\ee
is the free field part of the action, and
\be
S_{int} [\vf ,\vf^* ]\ =\ \int d^2 x\ V(\vf^* \vf ) \ ,
\ee
where $V(\cdot)$ is a positive definite polynomial describing the 
self-interaction of the field.

\subsection{Transition to noncommutative plane}
There exists a possibility to introduce in the two-dimensional Euclidean 
{\it coordinate} plane $P^{(2)}$ an additional Poisson structure. 
It is defined by the elementary bracket
\be
\{ x_i ,x_j \} \ =\ \varepsilon_{ij} \ ,\ i,j\ =\ 1,2\ ,    \e{poiss.br}
\ee
and extended by Leibniz rule to all smooth functions on $P^{(2)}$ (here
$\varepsilon_{ij}$ is the antisymmetric tensor, $\varepsilon_{12}=1$).
The brackets are invariant with respect to canonical transformations $x_i\ra
M_{ij}x_j+a_i$, where $M_{ij}$ is an unimodular matrix, $a_1,\,a_2$ are
arbitrary constants. In particular, the brackets are invariant with respect
to the two-dimensional group $E(2)$ of isometries of $P^{(2)}$ formed by
\be\ba{rll}
(i)&\mbox{rotations:} & x_1 \to x_1 \cos \f  + x_2 \sin \f,\qq
x_2 \to x_2\cos \f - x_1 \sin \f,\\[3mm]

(ii)& \mbox{translations:}& x_1 \to x_1 + a_1,\qq x_2 \to x_2 + a_2\ .\ea
\e{nc-plane.transf}
\ee

The noncommutative version $P^{(2)}_\ld$ of the plane is obtained 
by the  deformation (quantization)
of this Poisson structure. In the noncommutative approach we replace the
commuting parameters by the hermitian operators ${\hat x}_i$, $(i,j=1,2)$
satisfying commutation relations
\be
[{\hat x}_i ,{\hat x}_j ]\ =\ i\lambda^2 \varepsilon_{ij} 
\ ,\qq i,j\ =\ 1,2\ ,                          \e{op-comm}
\ee
where $\lambda$ is a positive constant of the dimension of {\it length}.
We realize the operators ${\hat x}_i$, $i,j=1,2$  in a suitable Fock space
${\cal F}$ introducing the annihilation and creation operators
\be
{\hat \alpha}\ =\ \frac{1}{\lambda\sqrt{2}}({\hat x}_1 + i{\hat x}_2 )\ ,\
{\hat \alpha}^\da \ =
\ \frac{1}{\lambda\sqrt{2}}({\hat x}_1 - i{\hat x}_2 )\ ,   \e{crea-annih}
\ee
and putting
\[
{\cal F} \ =\ \bigg\{ |n\rangle =
\frac{1}{\sqrt{n!}}{\hat \a}^{*n} |0\rangle ;\ n =0,1,\dots \bigg\} \ .
\]
Here $|0\rangle$ is a normalized state satisfying ${\hat \a}|0\rangle =0$.
This realization of operators ${\hat x}_i , i=1,2$ (or, equivalently of
${\hat \alpha}$ and ${\hat \alpha}^*$) corresponds to the unitary 
irreducible representation of the Heisenberg-Weyl group $H(2)$.

For all operators of the form
\be
{\hat f}\ =\ \frac{\ld^2}{(2\pi)^2} \int d^2 k\ {\tilde f}(k) e^{ik{\hat x}} 
\                                                      \e{QFTmom.rep}
\ee
(with a suitable smooth decreasing ${\tilde f}(k)$)
we introduce the integral (linear functional) $I_\ld[{\hat f}]$ as follows:
\be
I_\ld[{\hat f}]= {\rm Tr} {\hat f} = {\tilde f}(0) \ .  \e{NC-integr}
\ee
Here Tr denotes the trace in the Fock space and $k{\hat x}=k_1 {\hat x}_1
+k_2 {\hat x}_2$. The last equality in \r{NC-integr} 
indicates that $I_\ld[\cdot]$ is the noncommutative analog of the 
usual integral $I_0 [\cdot]$ on $P^{(2)}$.

The noncommutative analogs of field derivatives $\partial_i {\hat \vf}$,
$i=1,2$ are defined as
\be
\partial_i {\hat \vf} =  \varepsilon_{ij} \frac{i}{\lambda^2}
[{\hat x}_j ,{\hat \vf}]\ ,\qq i=1,2 \ .                          \e{NC-der}
\ee
They satisfy the Leibniz rule and reduce to the usual derivatives in the
commutative limit.

In the noncommutative case we define the Euclidean action of
self-in\-ter\-ac\-ting scalar noncommutative quantum field theory (NC-QFT) 
similarly as before: 
\be
S^{(\ld)}[{\hat \vf},{\hat \vf}^\da ]
=S_0^{(\ld)}[{\hat \vf},{\hat \vf}^\da ]
+S_{int}^{(\ld)}[{\hat \vf},{\hat \vf}^\da ]\ ,             \e{NCact-gen}
\ee
with the free part of the action
\be
S_0^{(\ld)}[{\hat \vf},{\hat \vf}^\da] 
= I_\ld[ (\partial_i {\hat \vf})^\da
(\partial_i {\hat \vf})\ 
+\ m^2 {\hat \vf}^\da {\hat \vf}] \ . \e{NC-free.action}
\ee
The interaction part $S_{int}$ of the action we shall discuss later (see 
\r{NC.interct} and below).

\subsection{Quantum mechanics with the Heisenberg-like relation for the 
coordinates on a quantum plane}

The existence of the fields $\wt\vf(k_1,k_2)$ in the momentum representation
with {\it commuting} variables $k_1,k_2$ implies that there exist commutative
operators of momentum components for a particle on the quantum plane
$P^{(2)}_\ld$. This is
indeed the case. The definition \r{NC-der} of the derivatives shows that the
momenta operators should be defined as follows (only in this subsection we 
use explicit dependence on the Planck constant $\h$, outside it we put 
$\h=1$):
\bn 
&&\Hp_i=\i\h\ld^{-2}\ve_{ij}\ad_{x_j}\ ,    \e{NC-mom}\\[3mm]
&&\ad_{x_i}\hA\equiv [\Hx_i,\hA]\qq \mbox{for any operator $\hA$}  
\en
and since the right-hand side of \r{op-comm} is a constant, 
we have 
\be 
[\Hp_1,\Hp_2]\sim[\ad_{x_2},\ad_{x_1}]=0\ .         \e{mom-comm} 
\ee  
Moreover, since the coordinate operators satisfy commutation relations of the
Heisenberg-Lie algebra, the form of the "\fr transform" \r{QFTmom.rep} shows
that the momentum variables $p_1,p_2$ play the role of parameters of the
corresponding Heisenberg-Weyl group.
Thus Quantum Mechanics which corresponds to NC-QFT is defined by the 
following commutation relations 
\bn 
&&[\Hx_1,\Hx_2]=\i\ld^2\ ,\qq [\Hp_1,\Hp_2]=0\ ,\nonumber\\[-1mm] 
&&\e{nc-phspace}\\[-1mm] 
&&[\Hx_i,\Hp_j]=\i\h\d_{ij}\ ,\qq i,j=1,2\ .\nonumber 
\en 
Introduction of the operators 
\be
\hX_i=\Hx_i+\frac{\ld^2}{2\h}\ve_{ij}\Hp_j\ ,    \e{comm-coord}
\ee
with the commutation relations
\bn 
&&[\hX_i,\Hp_j]=\i\h\d_{ij}\nonumber\\[-1mm] 
&&\e{commuting}\\[-1mm] 
&&[\hX_i,\hX_j]=[\Hp_i,\Hp_j]=0\ ,\nonumber 
\en 
shows that the quantum mechanical algebra of the phase space operators on the
considered quantum space (as well as the algebra in the model considered 
in \cite{DoplicherFR,Filk}) is the same as in the ordinary quantum mechanics.
This simple observation has two implications: 
\biz 
\item 
the very existence of noncommutativity, at least of the type considered in
this section, does not imply necessarily
UV-finiteness (contrary, for instance, to the case of Quantum Mechanics 
on a lattice); one 
needs some additional {\it dynamical} principle to achieve this aim; in 
particular, one may hope to construct a QFT Lagrangian in terms of 
noncommutative objects resulting in the essentially nonlocal action and
UV-finite QFT. 
\item 
the relation \r{comm-coord} show that for small values of the momentum the 
noncommutative coordinates are close to usual commutative coordinates: 
$$ 
\frac{\Hx_i}{\ld}=\frac{\hX_i}{\ld}-\frac12\ve_{ij}
\frac{\ld\Hp_j}{\h}\ , 
$$ 
so that if for a given state $\bra{\p}\Hp_i/\ld^{-1}\ket{\p}\ll1$, 
the two types of coordinates have close values. This 
remark may provide a basis for a more detailed explanation of the fact that 
particles with relatively low energy (\ie in any present experiment) do not 
feel the space-time noncommutativity. 
\eiz 

\subsection{Calculations by the method of operator symbols} 
The calculation of Green functions and other quantities in NC-QFT can be
carried out in simple and natural way by the use of operator symbols.  An
operator symbol (see, \eg \cite{BalazsJ} and refs. therein), as it is defined
in the ordinary Quantum Mechanics, is a function on the phase space which is
constructed by a definite rule from a given operator. Different rules for the
construction produce different symbols (\eg Weyl, normal symbols \etc). Sets of
such functions endowed with a so-called star-product form algebras which are
isomorphic to the initial operator algebra. Knowledge of explicit forms of
star-products ($\star$-products)  allows to make calculations in an easy and
short way.  In fact, transition to the momentum representation 
$\vf(\Hx)\ra\wt{\vf}(k)$ in eq. \r{QFTmom.rep} is the first step in the 
construction of the corresponding {\it Weyl symbol}. If, in addition, we make 
now inverse {\it ordinary} \fr transformation 
\be 
\vf_W(x)=\frac{\ld^2}{(2\pi)^2}\int\,d^2k\,\wt{\vf}(k)\er^{\i kx}\ ,\e{13} 
\ee 
we obtain just what is called the Weyl symbol $\vf_W(x)$ of the
operator $\vf(\Hx)$. Notice that the symbol is defined on the {\it
classical (commutative) analog} of the noncommutative space. Operator product 
corresponds to the $\star$-product in the set of the symbols: 
$\vf(\Hx)\p(\Hx)\lra\vf_W\star \p_W$. Thus, the Weyl symbols or their \fr
transforms (which plays the role of the momentum representation for a
field on the noncommutative plane $P^{(2)}_\ld$) 
are in one-to-one correspondence with the set
of fields (operators) $\vf(\Hx)$ on the noncommutative space. This
correspondence is based on the relation $\tr\exx{\i k_i\Hx_i}=2\pi\ld^{-2}
\d^{(2)}(k)$.
The trace of an operator $f(\Hx)$ is needed, in particular, for construction 
of the action. It is expressed via its Weyl symbol as follows:
$$ 
\tr\,f(\Hx)=\frac{1}{2\pi\ld^2}\int\,d^2x\,f_W(x)=I_\ld[f(\Hx)]\ . 
$$ 
Now any action for NC-QFT can be obtained from the corresponding classical 
action by the substitution of the ordinary point-wise function multiplication 
by the $\star$-product. For example,  
\bnn 
\tr\sum_i[\Hx_i,\vf(\Hx)]^2&=&\int\,d^2x\,\{x_i,\vf_W(x)\}_M\star 
\{x_i,\vf_W(x)\}_M\nbr 
&=&\int\,d^2x\,(\pa_i\vf_W\star\pa_i\vf_W)(x)\ , 
\enn 
where $\{\cdot,\cdot\}_M$ is the Moyal bracket: 
$$ 
\{\vf,\p\}_M\deff\frac{1}{\ld^2}\big(\vf\star\p-\p\star\vf\big)\ . 
$$ 
Equivalently, one may use the \fr transform $\wt\vf(p)$ of the Weyl symbol
(momentum representation). The existence of this field, depending on the 
commutative variables $p_1$ and $p_2$, corresponds to the commutativity of the
momentum operators (\cf \r{mom-comm}) of the considered system. 

The ``second quantization'' in the Euclidean case amounts to calculation of
the path integral over set of operator symbols which gives the generating
functional $Z[J]$ for Green functions
\be
Z[J]=N^{-1}\int\cD\vf_W(x)\exx{-S[\vf_W,J,\star]}    \ ,     \e{op.symb.pi}
\ee
where $S[\vf_W,J,\star]$ is the operator action \r{NCact-gen} on the 
noncommutative space expressed in terms of symbols, 
or, in other words, the usual classical action
in which the ordinary point-wise multiplication of the fields is substituted
by the star-product ($N$ is the normalization constant). 

The Weyl symbol has some special properties which makes it convenient for the
calculations. In particular:
\biz
\item[{\it i})]
The explicit form of the $\star$-product which makes the algebra of Weyl 
symbols isomorphic to the operator algebra is defined by the expression 
\bn 
\lefteqn{\left(\vf_W\star\p_W\right)(x)= 
\vf_W(x)\exp\left\{i\frac{\ld^2}{2} 
\stackrel{\leftarrow}{\partial}_i\ve^{ij} 
\stackrel{\rightarrow}{\partial}_j\right\}\p_W(x)}   \nonumber\\ 
&=&\sum_{m=0}^{\infty}\frac{1}{m!}\left(\frac{i\ld^2}{2}\right)^m 
\ve^{i_1j_1}\cdots\ve^{i_mj_m} 
\left(\partial_{i_1}\cdots\partial_{i_m}\vf_W\right) 
\left(\partial_{j_1}\cdots\partial_{j_m}\p_W\right) \nonumber\\ 
&=&\vf_W(x)\p_W(x)+{\cal O}(\eta) \ .              \e{Weyl-star} 
\en 
This immediately shows that a quadratic term in the NC-QFT action, 
written in terms of the Weyl symbols, has the same form as that on 
the classical space:
\bn
\int\,d^2x \left(\vf_W\star\p_W\right)(x)&=&
\int\,d^2x\vf_W(x)\exp\left\{-i\frac{\ld^2}{2} 
\stackrel{\rightarrow}{\partial}_i\ve^{ij} 
\stackrel{\rightarrow}{\partial}_j\right\}\p_W(x)\nbr
&=&\int\,d^2x\vf_W(x)\p_W(x)                    \ ,\e{Weyl-quadr}
\en
because $\ve^{ij}$ is antisymmetric. Therefore, the free action of the NC-QFT
in terms of the Weyl symbols has the same form as usual QFT on commutative
space. Higher order (interaction) terms contain non-locality, but Filk's
analysis in \cite{Filk} shows that they do not remove UV-divergences.
\item[{\it ii})]
One more property of the Weyl symbols is their nice behaviour 
with respect to linear canonical transformations: if one considers 
transformations (\cf \r{nc-plane.transf}) 
$$ 
\Hx'=M_{ij}\Hx_j+b_i\ , 
$$ 
the corresponding Weyl symbol transforms as follows 
$$ 
\vf_W'(x)=\vf_W(Mx+b) \ ,
$$ 
\ie it is transformed as an ordinary scalar field. This essentially 
simplify study of invariance properties of NC-QFT. 
\eiz

When considering the Green functions $G_W(x,y)\equiv\<\vf_W(x)\vf_W(y)\>$, 
one should take into
account that a value of an operator symbol at a point on the classical
counterpart of a noncommutative space has not direct physical and
even mathematical meaning. Only the total function can be considered as a
symbol and defines the corresponding operator. Thus the function $G_W(x,y)$
has the meaning of an operator symbol acting in the direct product 
$\cH\otimes\cH$ of
two copies of the Hilbert space in which a representation of the coordinate
algebra is realized. 
Now let us remember that in the standard QFT, the points of a commutative
space (labelled by values of the coordinates $x_1,x_2$) are considered to
``enumerate'' different degrees of freedom of the field system. In
noncommutative geometry there are no points anymore but there are states in
the Hilbert space of representations of a coordinate algebra instead. Thus we
must consider a quantum field in NC-QFT as a map from set of states on the
corresponding noncommutative space into the algebra of secondary quantized
operators, so that the physically meaningful object 
in NC-QFT is the mean values of the
field operators: $\bra{\Psi}\hat\vf\ket{\Psi}\,,\ \ket{\Psi}\in\cH$. Of
course, we can choose any complete set of states, but for clear physical
interpretation and comparing with the commutative limit, the set should
satisfy the following requirements:
\biz
\item[{\it i})] the states must be localized in space-time;
\item[{\it ii})] as the parameter of noncommutativity $\ld$ goes to zero, the
states must shrink to a point.
\eiz

This consideration shows that in order to convert the Green function 
$G_W(x,y)$ into the physically 
meaningful object, we must average it over some localized state. 
States which correspond to optimal rotationally invariant localization around
the point $x=(x_1,x_2)$ of the plane are uniquely (up to a phase factor) given
as the (non-normalized) coherent states $\ket{\xi}$ for the operators 
\r{crea-annih}: $\ket{\xi}=\exx{\xi\hat{\a}^\da}\ket{0}$ ($\ket{0}$ is 
the vacuum state in the Fock space $\cF$; $\xi=(x_1+\i x_2)/(\sqrt{2}\ld)$).

It can be shown that
\be
\vf_N(x)\equiv\frac{\bra{\xi}\hat\vf\ket{\xi}}{\<\xi|\xi\>}=
\int\,d^2x'\,\om_\ld(x-x')\vf_W(x')\ ,        \e{sm.field}
\ee
with the smearing function 
$$
\om_\ld(x-x')=\frac{1}{\pi\ld^2}\exx{-\frac{(x-x')^2}{\ld^2}}\ .
$$
In fact, $\vf_N(x)$ is the normal symbol of the operator
$\hat\vf$. The physical Green function $G_\ld(x,y)$ is therefore given as 
\bn
G_\ld(x,y)&=&\<\vf_N(x)\vf_N(y)\>\nbr
&=&\int\,d^2x\,d^2y\,\om_\ld(x-x')\om_\ld(y-y')G_W(x',y')\ , \e{sm.correl}
\en
and it represents a quantum average of the true field functional 
$\vf_N(x)\vf_N(y)=
\bra{\xi}\hat\vf\ket{\xi}\bra{\z}\hat\vf\ket{\z}$ (where $\xi=(x_1+\i
x_2)/\ld\sqrt{2}$, $\z=(y_1+\i y_2)/\ld\sqrt{2}$). Similarly, any higher
Green functions $G_\ld(x_1,...,x_n)$ are obtained by smearing the
corresponding Green functions  $G_W(x_1,...,x_n)$. 

A few remarks are in order:
\biz
\item[1)]
The formal Green functions $G_W(x_1,...,x_n)$ are, as a matter of rule, 
singular if some arguments coincide. 
However, the physical Green functions $G_\ld(x_1,...,x_n)$  are
regular due to {\it intrinsic} effective smearing induced by the
noncommutativity of the coordinates. We would like to stress that this is not
any artificial external smearing, and that no better localized Green functions 
as $G_\ld(x_1,...,x_n)$ can be constructed on the noncommutative plane. 
We illustrate this assertion on the example of a free field Green function. 
In this case the formal Green function 
$G^{(0)}_W(x,y)=\<\vf_W(x)\vf_W(y)\>_0$ is given by the standard formula
\be
G^{(0)}_W(x,y)=\frac{1}{(2\pi)^2}\int\,d^2k\,
\frac{\er^{\i k(x-y)}}{k^2+m^2}\ .                     \e{free.weyl.corr}
\ee
According to \r{sm.correl} the corresponding physical Green function 
$G^{(0)}_\ld =\<\vf_N(x)\vf_N(y)\>_0$ can be straightforwardly 
calculated with the result 
\be
G^{(0)}_\ld (x,y)=\frac{1}{(2\pi)^2}\int\,d^2k\,
\frac{\er^{\i k(x-y)-\ld^2k^2/2}}{k^2+m^2}\ .   \e{free.smeared.corr}
\ee
This can be easily derived by use of the normal symbols. In this case
the star-product for normal symbols has the form
\be
\vf_N(\bxi,\xi)\star\f_N(\bxi,\xi)=\vf_N(\bxi,\xi)\exx{\ld^2
\stackrel{\leftarrow}{\partial}_\xi
\stackrel{\rightarrow}{\partial}_\bxi}\f_N(\bxi,\xi) \ ,     \e{star.normal}
\ee
The free action in terms of the normal symbols takes the form
\bn
S_0^{(N)}&=&\int\,d^2\xi\, \bigg[\pa_i\vf_N(\bxi,\xi)
\exx{\ld^2\stackrel{\leftarrow}{\partial}_\xi
\stackrel{\rightarrow}{\partial}_\bxi}\pa_i\vf_N(\bxi,\xi)\nbr
&&+m^2\vf_N(\bxi,\xi)
\exx{\ld^2\stackrel{\leftarrow}{\partial}_\xi
\stackrel{\rightarrow}{\partial}_\bxi}\vf_N(\bxi,\xi)\bigg]\nbr
&=&\frac{1}{(2\pi)^2}\int\,d^2\k\, \wt\vf_N(-\k)\big(2\bar\k\k+m^2\big)
\er^{\ld^2\bar\k\k} \wt\vf_N(\k)\ .           \e{free.act.norm}
\en
Here $\wt\vf_N(\k)$ is the \fr transform of the normal symbol
$$
\vf_N(\bxi,\xi)=\frac{1}{(2\pi)^2}\int\,d^2\k\,
\er^{\i(\k\bxi+\bar\k\xi)}\wt\vf_N(\k)\ ,
$$
and $\k=\ld(k_1+\i k_2)/\sqrt{2}$. 

Whereas $G^{(0)}_W(x,y)$ is logarithmically divergent for $x\ra y$, the 
physical Green function is finite 
\be
\left|G^{(0)}_\ld(x,y)\right|\leq G^{(0)}_\ld(x,x)
=\frac{1}{(2\pi)^2}\int\,d^2k\,
\frac{\er^{-\ld^2k^2/2}}{k^2+m^2}\ ,                    \e{fin.smeared.corr}
\ee
depending only on a dimensionless parameter $a=\ld m$ characterizing the 
non-com\-mu\-ta\-ti\-vi\-ty.

\item[2)]
If the interaction is switched on, there naturally appears the problem 
of a perturbative determination of the full Green function 
$G_\ld =\<\vf_N(x)\vf_N(y)\>$. Within the perturbation theory the 
problem is reduced to the calculations of free field averages of the 
type $\<\vf_N(x)\vf_N(y)S_{int}^n\>_0$. However, now the problem 
of a noncommutative generalization of the interaction term arises. 
If we choose as a commutative prototype the $(\vf^*\vf)^2$-interaction, 
the most direct noncommutative generalization is
\bn
S_{int}^\ld[\hat\vf,\hat\vf^\da]&=&
gI[\hat\vf^\da\hat\vf\hat\vf^\da\hat\vf]\nbr
&=&g\int\,d^2x\,\vf^*_N(x)\star\vf_N(x)\star
\vf^*_N(x)\star\vf_N(x)\ .                            \e{NC.interct}
\en
This action produces vertices containing factors $\er^{\ld^2k^2/2}$  on each
leg with the momentum $k_i,\ i=1,2,3,4$, plus additional phase  factors
$\exx{\pm\i \ld^2(k_1\times k_2+k_3\times k_4)/2}$  (here $k\times
p\deff\ve_{ij}k_ip_j$). The Gaussian factor $\er^{-\ld^2k^2/2}$ from the
propagators are cancelled in Feynman diagrams and the UV-divergences appear.
This confirm Filk's analysis \cite{Filk} in terms of Weyl symbols. 
Of course, calculations with different types of
operator symbols, being different at intermediate steps, give the same
physical results. Notice, however, that the normal symbols of the field
operators on the noncommutative plane have much more clear physical
interpretation since they 
are related (in fact, equal) to mean values over localized coherent states.

\item[3)]
However, this is not the only possibility. Insisting only 
on a commutative limit
condition $\lim_{\ld\ra0}S_{int}^\ld[\hat\vf,\hat\vf^\da]=
S_{int}[\vf,\vf^*]$, the
integrand in the noncommutative integral 
$I[\hat\vf^\da\hat\vf\hat\vf^\da\hat\vf]$ is defined up
to the {\it operator ordering}. There is no problem to modify the operator
ordering of the generators $\Hx_1$ and $\Hx_2$ in the integrand
$\hat\vf^\da\hat\vf\hat\vf^\da\hat\vf$ 
in such a way that the vertices will not contain the
exponential factors $\exx{\ld^2k_i^2/2}$ on legs. 
For example, one can use the normal symbols for the construction of the free 
action but the Weyl symbols for the interaction part. The resulting action will
lead to UV-regular Feynman diagrams. However, besides this pragmatic point of
view, we have not been able to find any deeper principle preferring such
different ordering.
\eiz

\subsection{Symmetry transformations on the quantum plane\e{stqp}}
Some subgroup of the group of the canonical transformations of the commutation
relations for the coordinate operators can be considered as a group of
space-time symmetry for NC-QFT. As we discussed in the preceding subsection,
the degrees of freedom of NC-QFT correspond to a set of localized (\eg
coherent) states. Thus there appears the natural question  about behaviour of
such states under the quantum space-time symmetry transformations.

The fact that a linear transformations preserves commutation relations for a
set of some operators means that the latter are tensor operators. It is worth
to separate the case of commuting and noncommuting operators:
\biz
\item[1)]{\it A set of commutative operators}. 
{}For the general linear transformation of commutative operators
$x_i\ra x'_i=M_{ij}x_j+b_i$, where $M_{ij},\ b_i$ are ordinary 
$c$-number group parameters, the
vector $|\p_x\>$ remains an eigenvector of the transformed operator $x'$ but
with shifted eigenvalue $x'_i=M_{ij}x_j+b_i$. 
\item[2)]{\it A set\e{outlining} of noncommutative operators: 
tensor operator}. 
A tensor operator $\hA_i$ acting in some Hilbert space $\cH$, has, by the 
definition, the property
\be
\hA'_i\equiv M_{ij}(g)\hA_j=\hU(g)\hA_i\hU^{-1}(g)\ ,            \e{ad3}
\ee
where $M_{ij}(g)\ (i,j=1,...,d)$ is a matrix finite-dimensional
representation 
of a Lie group $G$, $g\in G$ and $U(g)$ is a unitary operator in the Hilbert 
space $\cH$. 
In general, the components $\hA_i\ (i=1,...,d)$ of a tensor operator 
do not commute with each other. Consider an eigenvector $\ket{\ld}_A$ of one 
component, say $\hA_d$, of the tensor operator. After the transformation, the
eigenvector $\ket{\ld}_{A'}$ of the transformed component $A'_d$ is related to
$\ket{\ld}_A$ by the operator $U(g)$:
\be
\ket{\ld}_{A'}=\hU(g)\ket{\ld}_A=
\sum_{\ld'}{}_A\!\bra{\ld}\hU(g)\ket{\ld'}_A^*
\ket{\ld'}_A\ .                                        \e{ad4}
\ee
Of course, if we transform both operators (as in \r{ad3}) {\it and} states of
the system: $\ket{\p}\ra\hU(g)\ket{\p}$, nothing changes (equal shifts of 
both a system and measurement devices lead to the same values of
measurements). 
To study behaviour of a system under the group transformations, we should
transform either observables (operators) or states of the system. Then
considering the action of transformed component $\hA'_d$ on the initial
eigenstate $\ket{\ld}_A$, we obtain
\bn
\hA'_d\ket{\ld}_A&=&\hU(g)\hA_d\hU^{-1}(g)\ket{\ld}_A\nbr
&=&\sum_{\ld'}\ld'\,{}_A\!\bra{\ld'}
\hU(g)\ket{\ld}_A\ket{\ld'}_{A'}\ .\e{ad5}
\en  
\eiz

Let us apply this consideration (well known in the standard Quantum Mechanics) 
to the examples of Euclidean and pseudo-Euclidean quantum planes. 
While UV-behaviour in these cases are the same, their
properties with respect to the symmetry transformations are quite different. 

We shall consider only homogeneous part of the transformations. In the case of
the Euclidean plane, these are rotations \r{nc-plane.transf} (one-dimensional 
subgroup  of the group $Sp(2)\sim SL(2,\R)$ of the canonical transformations).
The corresponding creation and annihilation operators \r{crea-annih} are
transformed separately
$$
\hat\a\ra\er^{\i\f}\hat\a\ ,\qq\hat\a^\da\ra\er^{-\i\f}\hat\a^\da\ ,
$$
so that the corresponding localized (coherent) states $\ket{\xi}$ are
transformed in very simple way:
\be
\ket{\xi}\lra\ket{\er^{\i\f}\xi}\ .         \e{coh-tr}
\ee
Thus the localized coherent states are transformed in the simple and
physically transparent way. On the contrary, coordinate eigenstates
are transformed non-locally according to \r{ad5}. Indeed, the coordinates are
transformed under Euclidean rotations by the formula
\bn
\Hx_1&\ra&\Hx'_1=(\cos\f)\, \Hx_1+
(\sin\f)\, \Hx_2=\hU_\f\Hx_1\hU_\f^{-1}\ ,\nonumber\\[-1mm]
                                       \e{NC-Euc-rot}\\[-1mm]
\Hx_2&\ra&\Hx'_2=-(\sin\f)\, \Hx_1+
(\cos\f)\, \Hx_2=\hU_\f\Hx_2\hU_\f^{-1}\ .\nonumber
\en
The explicit form of the operator $\hU_\eta$ is easily found and 
proves to be
$$
\hU_\f=\exx{-\frac{\i}{2}\f(\Hx_1^2+\Hx_2^2)}\ .
$$   
Formally this operator coincides with the evolution operator for a particle
in the harmonic potential, so that its kernel is well-known (see, \eg
\cite{FeynmanH}) 
\be
\<x_1'|\hU_\f|x_1\>=\frac{1}{\sqrt{2\pi{\i}\ld^2}}\sqrt{
\frac{1}{\sin\f}} \exx{\frac{1}{4\ld^2}
\frac{1}{\sin\f}\biggl[\bigg((x'_1)^2+x^2_1\bigg)\cos\f-2x'_1x_1\biggr]}
\ .                                             \e{NC-Euc-tens-tr}
\ee
Insertion of this kernel into the formulas \r{ad4}, \r{ad5} leads to the
nonlocal transformation of eigenstates of the operator $\Hx_1$ (eigenstates
of $\Hx_2$ are transformed quite similarly).
 
The situation is opposite in the case of pseudo-Euclidean (Minkowski) plane.
Now we have 
to use another subgroup of the canonical group $SL(2,\R)$: two-dimensional
Lorentz group $SO(1,1)$
\bn
x_0&\ra& x'_0=(\cosh\eta)\, x_0+(\sinh\eta)\, x_1\ ,\nonumber\\[-1mm]
\e{NC-nonE1}\\[-1mm]
x_1&\ra& x'_1=(\sinh\eta)\, x_0+(\cosh\eta)\, x_1\ ,\nonumber
\en
It is convenient to use the light-front variables
$$
x_\pm \ =\ \frac{1}{\sqrt{2}} (x_0 \pm x_1  ) \ .
$$
The boosts \r{NC-nonE1} now have the simple form
$$x_\pm \to e^{\pm \eta} x_\pm$$ 
($\eta$ has the meaning of rapidity). On the noncommutative plane the
coordinates satisfy the commutation relations
\be
[{\hat x}_+ ,{\hat x}_- ]\ =\ i\lambda^2 \ .
\ee
The corresponding annihilation and creation operators
\be
{\hat \alpha} \ =\ \frac{1}{\lambda \sqrt{2}} ({\hat x}_+ + i{\hat x}_- )
\ ,\ {\hat \alpha}^\da \ =\ \frac{1}{\lambda \sqrt{2}} ({\hat x}_+ -
i{\hat x}_- )\ ,
\ee
are transformed non-trivially
\bn
\hat\a&\ra&\hat\a_\eta=(\cosh\eta)\, \hat\a+
(\sinh\eta)\, \hat\a^\da=\hU_\eta\hat\a\hU_\eta^{-1}\ ,\nonumber\\[-1mm]
\e{NC-nonE2}\\[-1mm]
\hat\a^\da&\ra&\hat\a^\da_\eta=
(\sinh\eta)\, \hat\a+(\cosh\eta)\, \hat\a^\da
=\hU_\eta\hat\a^\da\hU_\eta^{-1}\ ,\nonumber
\en
The explicit form of the operator $\hU_\eta$ is easily found and proves to be
$$
\hU_\eta=\exx{-\frac12\eta\bigg((\hat\a^\da)^2-\hat\a^2\bigg)}\ .
$$
Thus now the corresponding localized coherent states are transformed as:
$$
\ket{\xi}\lra\ket{\xi_\eta}=\hU_\eta\ket{\xi}\ .
$$
The calculation of the matrix elements $\bra{\z}\hU_\eta\ket{\xi}=
\exx{-|\z|^2/2-|\xi|^2/2}\cU(\bzt,\xi)$ of the 
operator $\hU_\eta$ can be done, \eg with the help of the path integral
\bnn
\cU(\bzt,\xi)&=&\int\,\prod_\tau\frac{d\bz(\tau) dz(\tau)}{2\pi}
\exp\bigg\{\bar\z z(\eta)+\int_0^\eta\,d\tau\,
\big[-\bz(\tau)\dot z(\tau)\nbr
&&-\frac12\big(\bz^2(\tau)-z^2(\tau)\big)\big]\bigg\}\ .
\enn
This is Gaussian integral and, hence, its value is given by the integrand at
the extremum trajectory of the exponent with the boundary conditions:
$z(0)=\xi,\ \bz(\eta)=\bar\z$. The result is:
$$
\bra{\z}\hU_\eta\ket{\xi}=\exx{-\frac{|\z|^2}{2}-\frac{|\xi|^2}{2}+
\frac{\bar\z\xi}{\cosh\eta}-
\frac{\bar\z^2-\xi^2}{2}\tanh\eta}\ .
$$
Now to realize properties of the transformed state  
$\ket{\xi_\eta}=\hU_\eta\ket{\xi}$ we can calculate it in the coordinate
representation (either $x_+$ or $x_-$). For example,
$$
|\<x_+|\xi_\eta\>|^2\equiv\left|\<x_+|\hU_\eta\xi\>\right|^2
=\frac{1}{\sqrt{\pi}\ld\er^{-\eta}}
\exx{-\frac{\big(x_+-\ld\er^{-\eta}\xi_1\big)^2}{\ld^2\er^{-2\eta}}}\ 
$$
(here $\xi=(\xi_1+\i \xi_2)/\sqrt{2}$). This expression shows that
$\ket{\xi_\eta}$ is also localized state and with 
respect to the coordinate $x_+$ it is located around the point 
$\er^{-\eta}\sqrt{2}\ld\Ree{\xi}$ with the dispersion
$\big(\ld\er^{-\eta}\big)^2$ (while 
$\ket{\xi}$ is located around $\sqrt{2}\ld\Ree{\xi}$ with the dispersion
$\ld^2$).  Similarly, with respect to the coordinate $x_-$ the state 
$\ket{\xi_\eta}$ is located around the point 
$\er^{\eta}\sqrt{2}\ld\Imm{\xi}$ with the dispersion 
$\big(\ld\er^{\eta}\big)^2$. 

\section{Quantum field theory on a noncommutative cylinder}

The space-time (without dynamical gravitational fields) usually 
possesses some space-time symmetry described by a 
Lie group ${G}$ of isometries which transitively acts
on the space-time manifold. The group action of ${G}$ on a scalar field
$\vf(x)$ is defined by the equality
\be
T_g\vf(x)\deff\vf(g^{-1}x)\ ,\qq g\in{G}\,\qq x\in\cM\ .     \e{class.gr.act}
\ee
The right hand side of this definition can be considered as a function on
${G}\otimes\cM$. Let $\gg=\,Lie\,({G})$ denotes the Lie algebra of the group
in question, so that for any $\hX\in\gg$ there corresponds the group element
$g=\er^{t\hX}\in{G}$. Inserting this into \r{class.gr.act} we can assign to
any $\hX\in\gg$ the left-invariant vector field $\hX$ on $\cM$ defined by
\be
\hX\vf(x)\deff\lim_{t\ra0}\frac{1}{t}\left[\vf\left(\er^{-t\hX}x\right)
-\vf(x)\right]\ .                                          \e{left.vec.f}
\ee
In this way one obtains a representation of the Lie algebra $\gg$ (which, of
course, depends on the choice of $\cM$ and representation $T_g$ used in 
\r{class.gr.act}).

\subsection{Fields on noncommutative co-adjoint orbits of Lie groups}

Let $T_g$ be a unitary irreducible representation
of $G$ in the Hilbert space $\cH$. The corresponding Lie algebra generators
$\hX_1,...,\hX_n$ of $\gg$ satisfy the commutation relations
\be
[\hX_i,\hX_j]=\i c_{ij}^k \hX_k\ ,\qq i,j,k=1,...,n\ .  \e{Lie.alg.CR}
\ee
In a given representation these generators satisfy also the polynomial Casimir
operator relations:
\be
C_1(\hX_i)=\ld_1\one\ ,\qq\cdots,\qq C_d(\hX_i)=\ld_d\one\ , \e{Casimir.rel}
\ee
where $C_1(\hX_i)\,,\cdots,C_d(\hX_i)$ is the complete set of 
independent Casimir
operators, and the numbers $\ld_1,...,\ld_d$ characterize the representation
in question: $T=T_{\ld_1,...,\ld_d}$.

Let us now introduce the vector space $\gg^*$ dual to the vector space $\gg$
of the Lie algebra (\ie $\gg^*$ is the space of linear functionals on $\gg$:
$X\in\gg,\,Y\in\gg^*\,\ra\,\lgl Y|X\rgr\in\C$). 
The basis $Y^i$ of the space $\gg^*$ is defined by the standard relation: 
$\lgl Y^i|X_j\rgr=\d^i_{\ j}$. Now the generators $X_i$ can be considered as
coordinate functions on the dual space $\gg^*$. Suppose that the standard
commutative space-time $\cM$ can be identified with a maximal dimension
coadjoint orbit of the group ${G}$ in the space $\gg^*$ (considered as a usual
vector space). Such an orbit is defined by a set of values of the Casimir
polynomials
\be
C_1(x_i)=r_1\ ,\qq\cdots,\qq C_d(x_i)=r_d\ , \e{Casimir.orbit.rel}
\ee 
the Casimir polynomials depending now on {\it commutative} variables
$x_1,...,x_n$. Thus, functions on the commutative space-time (associated to
the orbit) are generated by
the commuting coordinates $x_1,...,x_n$ satisfying the conditions 
\r{Casimir.orbit.rel}.

The advantage of the interpretation of a space-time manifold as a coadjoint
orbit is that on the latter there exists the Lie-Kirillov-Kostant
symplectic structure (degenerated on the whole $\gg^*$ but well defined on
a coadjoint orbit), which leads to a Poisson brackets of the form
\be
\{x_i,x_j\}=C_{ij}^k x_k\ .                     \e{Lie-Kir-Konst}
\ee
The standard procedure of quantization, \ie replacement of the Poisson
brackets \r{Lie-Kir-Konst} by a commutator according to the rule 
$\{\cdot,\cdot\}\,\ra\,(1/\i\ld)[\cdot,\cdot]$, allows to associate to the
commutative manifold $\cM$ defined by \r{Casimir.orbit.rel}, the
noncommutative space $\cM_{NC}$. Functions on this noncommutative space are
generated by the operators $\Hx_i=\ld\hX_i$ with the commutation relations
\be
[\Hx_i,\Hx_j]=\i\ld C_{ij}^k \Hx_k\ ,                     \e{Lie-lambda}
\ee
satisfying the Casimir relations \r{Casimir.rel}, \ie generated by the Lie
algebra operators in the representation $T_{\ld_1\cdots\ld_d}$. The parameter
$\ld$ plays the role of the Planck constant and in the properly
defined limit one recovers the commutative limit \cite{Ber}.

Expressing the relevant differential operator $\D$ entering \r{clas.free.act}
in terms of the Poisson brackets \r{Lie-Kir-Konst}, it is then
straightforward to define its noncommutative analog by replacing brackets by
the commutator \r{Lie-lambda}. 

In the commutative case the fields on $\cM$ can be expanded as 
\be
\vf(x)=\sum_ka_{k_1\cdots k_n}x^{k_1}\cdots x^{k_n}\ ,   \e{field-power-ser}
\ee
and are transformed according \r{class.gr.act}. In the noncommutative case the
fields $\vf(\Hx)$ have analogous expansion
\be
\vf(\Hx)=\sum_ka_{k_1\cdots k_n}\Hx^{k_1}\cdots\Hx^{k_n}
\ ,                                 \e{NCfield-power-ser}
\ee
\ie they are operators in the Hilbert space $\cH$. Consequently, they
transforms according to the rule
\be
\vf(\Hx)\,\lra \,T_g\vf(\Hx)T_g^\da\ ,           \e{trans-NCfields}
\ee
as we discussed in the preceding section. 
Therefore, as a noncommutative generalization of the integral on $\cM$, one
can take a properly normalized trace
\be
I_\ld[\vf(\Hx)]=N(\ld)\tr\big(\vf(\Hx)\big)\ ,     \e{int-trace}
\ee
which is obviously ${G}$-invariant.
The noncommutative analog of the action $S[\vf,\vf^*]$ given by
\r{C-act} with the integral over $\cM$ is replaced by its
noncommutative analog \r{int-trace}, the field $\vf(x)$ being replaced by the
operators $\vf(\Hx)$.

It is well known that an orbit of a Lie group $G$ is in one-to-one
correspondence with a coset space $G/G_0$, where $G_0\subset G$ 
is the stability subgroup for a point on the orbit (see \eg \cite{BarutR}).
Then it is reasonable to associate to momenta degrees of freedom of the
corresponding NC-QFT the coset space $G/G_0$. 

Below we shall consider UV-properties of the field theory defined on a
noncommutative cylinder which can be considered as a quantization of a 
co-adjoint orbit of the Euclidean group $E(2)$ in the vector space $e^*(2)$
(dual to the Lie algebra $e(2)$).

\subsection{Fields on a commutative cylinder}
A standard cylinder with radius $\ro$ can be identified with the set of
points 
\be
C_\ro=\{(x,t),\ t\in\R,\ x=\ro\er^{\i\f},\ \ro=\const\}\ .   \e{com.cyl}
\ee
We shall interpret $C_\ro$ as a space-time manifold. Any function (field) 
on $C_\ro$ can be expanded as 
\bn
\vf(x,t)&=&\sum_{k=-\infty}^\infty\,a_k(t)\er^{\i k\f}\nbr
&=&a_0(t)+\sum_{k=1}^\infty \big[a_k^{(+)}(t)x^k_++a_k^{(-)}(t)x^k_-\big]
\ ,                                                        \e{cyl.exp}
\en
where $x_+=x,\ x_-=x^*,\ a_k^{(\pm)}(t)=a_{\pm k}(t)\ro^{-k},\ k=1,2,...$. As 
$I_0[\vf]$ we denote the usual integral on $C_\ro$
\be
I_0[\vf]\deff\frac{1}{2\pi\ro}\int_{C_\ro}\,dx\,dt\,\vf(x,t)
=\int_\R\,dt\,a_0(t)\ .                                \e{cyl-int}
\ee
The action for the field $\vf(x,t)$ we choose as follows
\be
S[\vf,\vf^*]=I_0\big[\vf^*\big(-\pa_t^2+\pa_\f^2-m^2\big)\vf
-V(\vf^*\vf)\big]\ .                                    \e{cyl.cl.act}
\ee
Since the cylinder $C_\ro$ can be interpreted as the co-adjoint orbit of the
two-dimensional Euclidean group $E(2)$, one can introduce on $C_\ro$
the (Lie-Kirillov-Kostant) Poisson bracket 
\be
\{\vf_1,\vf_2\}=\pad{\vf_1}{x_\mu}\pad{\vf_2}{x_\nu}
\{x_\mu,x_\nu\}\ ,                        \e{cPbr}
\ee
where the indices take values $\mu,\nu=0,+,-$ and we put $x_0=t$. 
The elementary brackets are given as 
\be
\{x_0,x_\pm\}=\pm\i x_\pm\ ,\qq \{x_+,x_-\}=0\ .           \e{elem.P.br}
\ee
These are exactly $e(2)$ Lie algebra relations. The point is that the function
$x_+x_-$ is central: $\{x_0,x_+x_-\}=\{x_\pm,x_+x_-\}=0$. Therefore, the
constraint $x_+x_-=\ro^2$ is consistent with the $e(2)$ Lie algebra structure
\r{cPbr}. It can be straightforwardly shown that 
\be
\pa^2_t\vf=\ro^2\{x_+,\{x_-,\vf\}\}\ ,\qq 
\pa^2_\f\vf=-\{x_0,\{x_0,\vf\}\}\ .                     \e{der-Pbr}
\ee
Therefore the free field action can be represented in terms of the Poisson 
brackets as
\be
S_0[\vf,\vf^*]=I_0\big[\vf^*\big(\Box+m^2\big)\vf\big]\ ,  \e{cyl.cl.act.Pbr}
\ee
where 
\be
\Box\vf=\{x_0,\{x_0,\vf\}\}+\ro^2\{x_+,\{x_-,\vf\}\}\ .     \e{dalamb.Pbr}
\ee
Inserting expansion \r{cyl.exp} into \r{cyl.cl.act} we obtain the free field
action in the form
\be
S_0[\vf,\vf^*]=\int_\R\,dt\,\sum_{k=-\infty}^\infty\big[\dot a_k^*(t)
\dot a_k(t)-(k^2+m^2)a_k^*(t)a_k(t)]\ .              \e{act.modes}
\ee
Similarly, $S_{int}[\vf,\vf^*]$ is a higher order polynomial in 
$a_k^*(t)$ and $a_k(t)$. 

Performing the \fr transform
\be
\wt a_k(\om)=\frac{1}{2\pi}\int_\R \,dt\,a_k(t)\er^{-\i \om t} \ ,  \e{en.fr}
\ee
we obtain the free field action in the diagonal form
\be
S_0[\vf,\vf^*]=\frac{1}{2\pi}\int\,d\om\,\sum_{k=-\infty}^\infty
\wt a_k^*(\om)(\om^2-k^2-m^2)\wt a_k(\om)]\ .              \e{act.modes.fr}
\ee
It follows then straightforwardly, that the free field Green function is 
\be
G_0(x',t';x,t)=\frac{1}{2\pi}\int_\R\,d\om\,\sum_{k=-\infty}^\infty
\frac{\er^{\i\om(t'-t)}}{\om^2-k^2-m^2+\i\ve}
\left(\frac{x'x^*}{\ro^2}\right)^k \ .                 \e{cl.cyl.Gr.f}
\ee
The Green function is diagonal in the energy-momentum representation, \ie 
\be
\wt G_0(\om,k)=\frac{1}{\om^2-k^2-m^2+\i\ve}\ .        \e{cl.cyl.Gr.f.en-mom}
\ee 
The vertex contribution in the momentum representation, \eg for 
the potential term $V=g(\vf^* \vf )^2$, has the form
\be
g\ \delta (\omega_1 -\omega_2 +\omega_3 -\omega_4 )\
\delta_{k_1 +k_3 ,k_2 +k_4}\ . \ \ \                \e{pp-add69}
\ee
Here, $(\omega_i ,k_i )$ with $i=1,3$, or $i=2,4$, correspond to 
two fields $\vf$, or $\vf^*$ respectively, in the vertex. This 
leads to a divergent tadpole diagram 
which, up to the interaction constant $g$, are given as
\bn
G_0(x,t;x,t)&=&\frac{1}{2\pi}\int_\R\,d\om\,\sum_{k=-\infty}^\infty
\frac{1}{\om^2-k^2-m^2+\i\ve}\nbr
&=&\sum_{k=0}^\infty\frac{1}{\sqrt{k^2+m^2}}=\infty \ . \e{cl.cyl.tadpole}
\en
This leads to UV-divergences in Feynman diagrams calculated within 
perturbation theory for the model defined by the action \r{cyl.cl.act}. 

\subsection{Field theory on a noncommutative cylinder}
The starting point is the replacement of commutative variables 
$x_0,x_\pm$ by the noncommutative ones $\Hx_0,\Hx_\pm$, satisfying 
the $e(2)$ Lie algebra commutation relations
\be
[\Hx_0,\Hx_\pm]=\pm\ld\Hx_\pm\ ,\qq [\Hx_+,\Hx_-]=0\ .      \e{NCcyl-coord}
\ee
The $e(2)$ possesses one series of infinite dimensional unitary 
representations $\pi_\ro$ in which 
\be
\Hx_+\Hx_-=\ro^2\ ,\qq \ro>0\ ,                         \e{NCcyl.Casim}
\ee
(to simplify notation we use the same symbol for generators and their 
representatives). This representation can be realized in the Hilbert space 
$\cL^2(S^1,d\f/2\pi)$ as follows
\be
\Hx_0=-\i\ld\pa_\f\ ,\qq \Hx_\pm=\ro\er^{\pm\i\f}\ .   \e{NCcyl-rep}
\ee
In the basis $\ket{n}=\er^{\i n\f}\,,\ n\in\Z$, the operator $\Hx_0$ is 
diagonal 
\be
\Hx_0\ket{n}=\ld n\ket{n}\ .                        \e{NCcyl-time}
\ee
Notice, that there exist a set of inequivalent representations 
$\ket{n}_\a=\er^{\i(n+\a)\f}\,,\ \ 0\leq\a<1$. For simplicity, we shall use
only the representation \r{NCcyl-time} which corresponds to $\a=0$.
The field $\hat\vf=\vf(\Hx_0,\Hx_\pm)$ is an operator in 
$\cL^2(S^1,d\f/2\pi)$ and we suppose that it possesses the expansion
\be
\hat\vf=a_0(\Ht)+\sum_{k=1}^\infty \big[a_k^{(+)}
(\Ht)\Hx^k_++a_k^{(-)}(\Ht)\Hx^k_-\big]\ ,      \e{NCcyl.exp}
\ee
The noncommutative integral $I_\ld[\vf]$ we define by
\be
I_\ld[\hat\vf]\deff\ld\tr\hat\vf=\ld\sum_{n\in\Z}\,a_0(\ld n)\ .\e{NCcyl-int}
\ee
We see that the usual measure on $\R$ in \r{cyl-int} is replaced by a 
uniform step measure in points $\ld n,\ n\in\Z$ with the height $\ld$. 

All dynamical information which is contained in the eigenvalues $a_k(\ld n)$ 
of operators $a_k(\Ht)$ is contained in the \fr transformed coefficients
\be
\wt a_k(\om)=\frac{1}{2\pi}\sum_{n\in\Z}\,a_k(\ld n)
\er^{-\i\om\ld n} \ ,                           \e{NCfr-tr}
\ee
defined on the interval $\om\in(-\pi/\ld,+\pi/\ld)$.

The free field action has the form similar to that in the commutative case:
\be
S_0[\hat\vf,\hat\vf^*]=
I_\ld\big[-\vf^*\big(\Box_\ld+m^2\big)\vf\big]\ ,  \e{NCcyl.cl.act.Pbr}
\ee
where the noncommutative analog $\Box_\ld$ of the operator  
$\Box$ is defined as (\cf \r{dalamb.Pbr})
\be
\Box_\ld\hat\vf=-\frac{1}{\ld^2}[x_0,[x_0,\vf]]+\frac{1}{\ld^2\ro^2}
[x_+,[x_-,\vf]]\ .     \e{NCdalamb.Pbr}
\ee
Taking $\hat\vf$ in the form
\be
\hat\vf=\sum_{k=-\infty}^\infty a_k(\Ht)\Hx^{(k)}\ ,    \e{NCfiel.exp}
\ee
(here $a_{\pm k}(\Ht)=a^{(\pm)}_k(t),\ \Hx^{(\pm k)}=\Hx^k_\pm,\ k=1,2...$),
we obtain the analog of the free field action \r{act.modes}, 
however, with a discrete time evolution
\bn
S_0[\hat\vf,\hat\vf^\da]&=&\ld\sum_{n,k}\bigg[-\frac{1}{\ld^2}a_k^\da(n\ld)
\big[a_k(n\ld+\ld)-2a_k(n\ld)+a_k(n\ld-\ld)\big]\nbr
&&-a_k^\da(n\ld)
(k^2+m^2)a_k(n\ld)\bigg]\ .              \e{NCact.modes}
\en
This action can be diagonalized by performing the \fr transformation 
(inverse to \r{NCfr-tr})
\be
a_k(\ld n)=\frac{\ld}{2\pi}\int_{-\pi/\ld}^{\pi/\ld}\,d\om\,\wt a_k(\om)
\er^{\i\om\ld n} \ ,                           \e{NCfr-tr-inverse}
\ee
The free field action on a noncommutative cylinder in the momentum 
representation takes the form
\be
S_0[\hat\vf,\hat\vf^*]=\frac{\ld}{2\pi}\int_{-\pi/\ld}^{\pi/\ld}\,d\om\,
\sum_{k=-\infty}^\infty \wt a_k^*(t)\big[\OM^2_\ld(\om)-k^2-m^2\big]
\wt a_k(\om)\ ,                              \e{NCact.modes.cyl}
\ee
where $\OM_\ld(\om)=\frac{2}{\ld}\sin\frac{\om\ld}{2}$. 
The corresponding free field Green function is
\be
G_0^{(\ld)}(x',t';x,t)=\frac{1}{2\pi}\int_{-\pi/\ld}^{\pi/\ld}\,d\om\,
\sum_{k=-\infty}^\infty
\frac{\er^{\i\om(t'-t)}}{\OM^2_\ld(\om)-k^2-m^2+\i\ve}
\left(\frac{x'x^*}{\ro^2}\right)^k \ ,                 \e{NCcl.cyl.Gr.f}
\ee
with the time variables taking discrete values: $t=\ld n,\ t'=\ld n'$. 
Notice, that the energy integral is carried out over {\it finite} interval.
In the energy-momentum representation the Green function is diagonal 
\be
\wt G_0^{(\ld)}(\om,k)=\frac{1}{\OM^2_\ld(\om)-k^2-m^2+\i\ve}
\ .                                          \e{NCcl.cyl.Gr.f.en-mom}
\ee
In the noncommutative case the interaction term we take in the form
$V=\frac{g}{4}:(\vf^* \vf )^2 :$, where the ordering $:\cdots :$
means that all expansion coefficients of fields $a_k (\Ht),a^\da_k (\Ht)$
are collected as a right (or equivalently left) factor. This
guarantees that there are no artificial time derivatives in the 
potential: all operators $a_k(\Ht),a^\da_k (\Ht)$ 
in the potential term appear at the same time $t=\lambda n$. The
vertex contribution has the same form \r{pp-add69} as in the commutative
case with the only difference that the symbol $\delta 
(\omega -\omega' )$ now refers to the periodic $\delta$-function
on the interval $[-\pi /\lambda ,+\pi /\lambda ]$. The tadpole
contribution is {\it finite}
\be
G_0(x,t;x,t)=\frac{1}{2\pi}\int_{-\pi/\ld}^{\pi/\ld}\,d\om\,
\sum_{k=-\infty}^\infty\frac{1}{\OM^2_\ld(\om)-k^2-m^2+\i\ve}<\infty
\ .                                                \e{NCcl.cyl.tadpole}
\ee
This can be seen most directly using the well-known summation formula 
(see \cite{GradshteynR}, p.36) 
$$
\sum_{k=-\infty}^\infty\frac{1}{x^2-k^2}=\frac{\pi}{x}\cot (x)\ .
$$
Thus,
\be
G_0(x,t;x,t)=\int_{0}^{\pi/\ld}\,d\om\,
\frac{\cot\left(\sqrt{\OM^2_\ld(\om)-m^2+\i\ve}\right)}
{\sqrt{\OM^2_\ld(\om)-m^2+\i\ve}}\ .             \e{NC.cyl.exp.tadpole}
\ee
The integrand in \r{NC.cyl.exp.tadpole} has one simple pole at 
$\om=\frac{2}{\ld}\arcsin\frac{\ld m}{2}$, which due to the 
$(-\i\ve)$-prescription leads to a finite contribution. The rest of integrand 
possesses the integrable singularities $(\om-\om_k)^{-1/2}$ at points 
\bnn
\om_k&=&\frac{2}{\ld}
\arcsin\left(\frac{\ld}{2}\sqrt{m^2+k^2\pi^2}\right)\ ,\nbr
&&k\ {\mbox{- integers}}\ ,\qq 
0<k\leq\frac{1}{\pi}\sqrt{4/\ld^2-m^2}\ .
\enn 
Since the integration range is 
restricted, this means that the tadpole integral is finite.

In the two-dimensional scalar field theory on commutative space the tadpole is
the only divergent contribution to Feynman diagrams. We have shown that
transition to the noncommutative cylinder leads to {\it UV-finite field
theory}. 

\section{ Quantum mechanics and quantum field theory on a noncommutative 
quantum plane with $E_q(2)$-symmetry}    

In this section we consider Quantum Mechanics induced by a quantum group 
structure.
Recall that in the case of ordinary Lie group $G$, the group structure defines
a unique symplectic structure on the cotangent bundle $T^*_G$ to the group
manifold $G$  and, hence, the corresponding
canonical quantization (via substitution of Poisson brackets by the
corresponding commutators). A similar construction with necessary
generalizations, can be carried out for Lie-Poisson groups,
which after the quantization procedure become quantum groups (see, \eg review
in \cite{ChaichianDbk} and refs. therein). 

Thus instead of starting from ordinary $E(2)$, we shall use below its quantum
version $E_q(2)$  \cite{VaksmanK,SchuppWZ,BonecciCGST}. Though in the case of
quantum groups and corresponding quantum homogeneous spaces (definition of the
latter see, \eg in \cite{BonecciCGST}) group parameters (coordinates) become
noncommutative,  the general scheme of quantization still can be applied. The
role of momentum operators is now attributed the $q$-deformed left- (or right-)
invariant generalizations of vector fields. Thus the Planck constant $\h$
enters, as usual, the commutation relations for momenta and coordinates, while
the group deformation parameter $q$ governs non-trivial coordinate-coordinate
and momentum-momentum commutation relations.  Therefore, first of all we have
to construct possible representations of this {\it combined}  $q$-deformed
algebra of noncommuting coordinates and momenta.  For the particular case which
we consider in this paper ($q$-deformed quantum Euclidean plane $P^{(2)}_q$)
this is not a very complicated problem.

At the next step we proceed to the construction of NC-QFT on the $q$-plane 
$P^{(2)}_q$ and to the study of its UV-properties.

\subsection{Algebra of coordinates and momenta on $P^{(2)}_q$ and its
representations}
In order to simplify notation, we shall omit, in what follows, the sign
``hat'' over operators.

We start from the quantum group $E_q(2)$ generated by elements $\bv,\ v,\
\bt,\ t$ with the defining relations \cite{VaksmanK} 
\be\ba{lllll}
\bv v = v\bv = 1\ ,& \qq &t\bt = q^2 \bt t\ ,& \\[3mm]
vt= q^2tv\ ,& \qq &\bv t = q^{-2}t\bv &\qq& q\in\R\ .\ea       \e{2.1}
\ee
Other commutation relations follow from the involution: 
$v^\da =\bv,\ t^\da=\bt$. 
The comultiplication has the form
\begin{equation}
\begin{array}{cc}
\D v = v \otimes v\ , & \D\bv = 
\bv \otimes \bv\ , \\[2mm]
\D t = v \otimes t + t \otimes \one\ , &
\D\bt = \bv \otimes \bt + \bt \otimes {\one}\ .
\end{array}                                            \label{2.2}
\end{equation}
The explicit form of other basic maps for $E_q(2)$ (antipode, counity) will
not be used in what follows. 

The unitary element $v$ can be parameterized with the help of the symmetric
element $\th$:
\be\ba{lcl}
v=\er^{\i\th}\ ,&\qq&\th^\da=\th\ ,\\[2mm]
\ &\ &\D\th=\th\otimes\one+\one \otimes \th\ .\ea      \e{2.4}
\ee
The corresponding quantum universal enveloping algebra (QUEA) $\cU_qe(2)$ is
generated by the elements $J,\ \bar T,\ T$ which are dual to the generators
$\th,\ \bt,\ t$ of the algebra $E_q(2)$ and, as a result of the duality, 
satisfy the following commutation relations 
\be\ba{lll}
[J,T]=\i T\ ,&\qq& [J,\bar T]=-\i \bar T\ ,\\[2mm]
T\bar T=q^2\bar T T&&\ea                                 \e{2.6}
\ee
(comultiplication and the other basic maps are also defined by the duality).

The left action of elements from QUEA $\cU_q L$ of an arbitrary 
Lie algebra $L$ on
elements of the corresponding quantum group $G_q$ is defined by the
expressions (see \cite{ChaichianDbk} and refs. therein)
\be
\ell (X)f=(\id\otimes X)\cdot\D f\equiv\sum_if^i_{(1)}
\lgl X,f^i_{(2)}\rgr\ ,                                  \e{2.8}
\ee
or
\be
\ld (X)f=(S(X)\otimes \id)\cdot\D f\equiv\sum_i\lgl S(X),f^i_{(1)}\rgr
f^i_{(2)}\ ,                                  \e{2.8a}
\ee
where $X\in\cU_q L,\ f,f^i_{(1,2)}\in G_q$, $\lgl\cdot,\cdot\rgr$ denotes the
duality contraction, $S(X)$ is antipode and where the comultiplication 
in $G_q$ is presented in the
form $\D f= \sum_if^i_{(1)}\otimes f^i_{(2)}$. An explicit calculation of this
left action in the case of $E_q(2)$ shows that the operators
$\bar T,T\in\cU_qe(2)$ act on elements of $E_q(2)$ generated by $\bt,\ t$ 
exactly in the same way as the q-deformed derivatives $\bpaq,\ \paq$
\cite{WessZ}. 
In fact, the elements $\bt,\ t$ generates $q$-deformed analog of
homogeneous space $P^{(2)}_q= E_q(2)/U_q(1)$, \ie algebra of functions on
quantum Euclidean plane \cite{BonecciCGST}. We shall denote elements of the
algebra $P^{(2)}_q$ by $\bz,\ z$ to distinguish them from elements $\bt,
\ t$ of the algebra  $E_q(2)$. 

The elements $\bz,\ z$ and $\bpaq,\ \paq$ defines the q-deformed algebra of
functions on $P^{(2)}_q$ together with the $q$-deformed left-invariant vector 
fields (derivatives). Its defining relations read as
\bn
&&z\bz=q^2\bz z\ ,\qq \pa_q\bpa_q=q^2\bpa_q\pa_q\nbr
&&\pa_q z=1+q^{-2}z\pa_q\ ,\qq\bpa_q \bz=1+q^{2}\bz\bpa_q\ ,\e{2.9}\\[3mm]
&&\bpa_q z=q^{2}z\bpa_q\ ,\qq\pa_q \bz=q^{-2}\bz\pa_q\ ,\nonumber
\en
(the commutation relations for the $q$-derivatives is just rewritten
commutation relations for $\bar T,\ T$ 
\r{2.6} and
those for the $q$-derivatives and coordinates are derived from \r{2.8}). If we
put $q=1$ and define $p=-\i\h\pa,\ \bar p=-\i\h\bpa$, the relations \r{2.9}
become the usual canonical commutation relations for a particle in 
two-dimensional space.
The requirement of consistency with antipode dictates the following
conjugation rule for the $q$-derivatives \cite{SchuppWZ}
\be
\pa_q^\da=-q^2\bpa_q\ ,\qq\bpa_q^\da=-q^{-2}\pa_q\ .        \e{2.10}
\ee

We consider the relations \r{2.9} as a $q$-deformation of the 
canonical commutation relations and
is going to construct their representation in a Hilbert space.

To this aim let us introduce the operators $N$ and $\bN$ defined by the
relations
\be
[N;q^{-2}]=z\pa_q\ ,\qq [\bN;q^{2}]=\bz\bpa_q\ ,          \e{2.11}
\ee
\be
[X;q^{\a}]\equiv\frac{q^{\a X}-1}{q^{\a}-1}\ .  \e{q-number}
\ee
These operators have simple commutation relations 
\bn
&&q^{\a N}z=q^\a zq^{\a N}\ ,\qq q^{\a \bN}\bz=q^\a \bz q^{\a \bN}
\ ,\nonumber\\[-1mm]
&&          \e{2.15} \\[-1mm]
&&q^{\a N}\pa_q=q^{-\a}\pa_q q^{\a N}\ ,\qq
 q^{\a\bN}\bpa_q=q^{-\a}\bpa_q q^{\a\bN}\ .\nonumber
\en  
Using \r{2.10} and $z^\da=\bz$, we find
\be
N^\da=-\bN-1\ ,\qq \bN^\da=-N-1\ .                \e{2.16}
\ee 
With the help of operators $N$ and $\bar N$ we can construct two operators, a 
hermitian $J$ and an antihermitian $D$ given as
\be
J=(\bN-N)\ ,\qq D=N+\bar N+1\ ,               \e{2.19a}
\ee
which correspond to rotations and dilatations on the quantum plane. Their
action on polynomial functions in $z$ and $\bar z$ is given as
\bn
q^J\,z^n\bar z^m&=&q^{m-n}z^n\bar z^m\ ,\nonumber\\[-1mm]
&&                          \e{new.add1}\\[-1mm]
q^D\,z^n\bar z^m&=&q^{m+n+1}z^n\bar z^m\ .\nonumber
\en
We stress that only $J$ represents a well-defined $\cU_qe(2)$ action on a
quantum plane, when supplementing $J$ by $T=\paq$ and $\bar T=\bpaq$:
\bn
T\,z^n\bar z^m&=&[n,q^{-2}]z^{n-1}\bar z^m\ ,\nonumber\\[-1mm]
&&                          \e{new.add2}\\[-1mm]
\bar T\,z^n\bar z^m&=&q^{2n}[n,q^{-2}]z^n\bar z^{m-1}\ .\nonumber
\en
The dilatation operator $D$, however, corresponds to a well defined
transformations within the $q$-deformed Heisenberg algebra \r{2.9} (in \r{2.9}
the operators $q^{-2N}=z\paq$ and $q^{2\bar N}=\bar z\bpaq$ appear
independently).

The operators $q^{2\bN},\ q^{2N}$ allow to construct commuting pairs of
conjugate operators:
\bn
&&\bar Z= q^{N-\bN}\bz\ ,\qq Z= zq^{N-\bN}\ ,\nonumber\\[-1mm]
&&\e{2.20}\\[-1mm]
&&\bar P= qq^{-(N-\bN)}\bpa_q\ ,\qq P= -q^{-1}\pa_qq^{-(N-\bN)}\ ,\nonumber
\en
with the commutation relations
\bn
&&P\bar Z=\bar Z P\ ,\qq \bar Z Z=Z\bar Z\ ,\qq
 ZP=1+q^{2}PZ \ ,\nonumber\\[-1mm]
&&\e{2.22}\\[-1mm]
&&\bar P Z=Z\bar P\ ,\qq \bar P P=P\bar P\ ,\qq
\bar P\bar Z=1+q^{2}\bar Z\bar P\ .\nonumber
\en
If we were given only the algebra of the operators $\bz,\ z,\ \bpa_q,\ \pa_q$,
we would reasonably name the commuting operators $\bar Z,\ Z$ by coordinates
and $\bar P,\ P$ by the corresponding lattice momenta and then deal with two
independent (commuting with each other) one-dimensional algebras on the 
$q$-lattice (for a one-dimensional Quantum Mechanics on $q$-lattice see, \eg
\cite{ChaichianDbk} and refs therein). 
However, fields in NC-QFT depend on noncommutative ($q$-commuting)
coordinates $\bz,\ z$ which are more suitable to trace a
result of coaction by $E_q(2)$. We have found convenient to use the hermitian
and unitary combination of the coordinate operators:
\be
r^2\equiv\bz z\ \ \ \mbox{(hermitian)},\qq u\equiv\bz z^{-1}
\ \ \ \mbox{(unitary)},                                 \e{2.37a}
\ee
together with $q^{2(\bN-N)},\ q^{2(\bN+N+1)}$ as a basic set of the phase
space operators. The commutation relations for this set of operators
read as
\be\ba{lll}
[q^{2(\bN-N)},r^2]=0\ ,&\qq &[q^{2(\bN+N+1)},u]=0\ ,\\[3mm]
r^2u=q^4u r^2\ , & \qq & [q^{2(\bN-N)},q^{2(\bN+N+1)}]=0\ ,\\[3mm]
q^{2(\bN-N)}u=q^4u q^{2(\bN-N)}\ , & \qq &
q^{2(\bN+N+1)}r^2=q^4r^2q^{2(\bN+N+1)}\ ,\ea   \e{2.38}
\ee

Now we are ready to construct a representation of this algebra in the space
$\ell^2$ (\ie infinite dimensional matrix representation):
\bn
r^2\mid n,m\rangle_{r_0,l_0}  & = &  
r^2_0q^{4n}\mid n,m\rangle_{r_0,l_0} \nbr
q^{2(\bN-N)}\mid n,m\rangle_{r_0,l_0} & = 
& l_0q^{4m}\mid n,m\rangle_{r_0,l_0}\ .       \label{2.39}
\en
\bn
u\mid n,m\rangle_{r_0,l_0} & = & \mid n+1,m+1\rangle_{r_0,l_0}\ , \nbr
q^{2(\bN+N+1)}\mid n,m\rangle_{r_0,l_0}& = &\mid n+1,m\rangle_{r_0,l_0}\ .
                                          \label{2.40}
\en
The constants $r_0$ and $l_0$ mark different representations and from
the eigenvalues of $r^2$ and $q^{2(\bN-N)}$ it follows that in the ranges
$[r_0,q^4 r_0)$ and $[l_0,q^4l_0)$ the representations are inequivalent.
The matrices $r^2,q^{2(\bN-N)}$ are hermitian and $\Phi,q^{2(\bN+N+1)}$ 
are unitary with respect to the scalar product defined by
$$
{}_{r_0,l_0}\langle
n,m\mid n',m'\rangle_{r_0,l_0}=\delta_{nn'}\delta_{mm'}\ .      $$

Thus we have obtained that states of particle on a quantum plane are
characterized by discrete values of its radius-vector and discrete values of
the operator $q^{2(\bN-N)}$ which is obviously related to deformation of the
angular momentum operator. Indeed, from \r{2.19a} we conclude that the
operator 
\be
J_q\equiv[\bN-N;q^2]=\frac{q^{2(\bN-N)}-1}{q^{2}-1}\ ,   \e{2.41}
\ee
(which differs from $q^{2(\bN-N)}$ by multiplication and shifting by the
constants) in the continuum limit $q\ra1$ becomes the ordinary angular 
momentum operator. Therefore it is natural to consider $J_q$ as an appropriate
deformation of the latter. Of course, discreteness of values of an angular
momentum operator is not peculiar feature of $q$-deformed systems but general
property of all quantum systems.

It is worth to mention that the coordinate algebra of the quantum plane
$P_q^{(2)}$ is closely related to the coordinate algebra \r{NCcyl-coord} of
the quantum cylinder. Indeed, if operators $x_0,x_+$ obey the commutation
relations \r{NCcyl-coord}, the operators $r^2=\er^{x_0},\ u=x_+$ serve as the
quantum coordinates \r{2.37a}, \r{2.38} on the quantum plane
$P_q^{(2)}$. This correspondence of the algebras is a non-commutative analog
of the well-known exponential map of a cylinder to a plane with one punctured
point. One can check that the exponentiation of the operator 
$x_0:\ x_0\ra r^2=\er^{x_0}$, does not change properties of the algebra
representations. However, this relation between the algebras does not mean
that properties of NC-QFT constructed on the quantum cylinder and on
$P_q^{(2)}$ are the same. Different symmetry properties and different
structure of phase spaces lead to quite different field theories on these
closely related non-commutative spaces. From technical point of view, the
corresponding symmetry principles induce different constructions of the
derivatives (the commutators in the case of the cylinder (\cf
\r{NCdalamb.Pbr}) and $q$-derivatives on the $q$-plane) and, hence, different
form of free actions in the two cases. While NC-QFT on the non-commutative
cylinder has no divergences, the theory on the $q$-plane $P_q^{(2)}$, as we
shall see later, has the usual UV-divergences (similar to the case of the
field theory on a commutative plane).

In the case of quantum group symmetries there appears a new type of 
noncommutativity: parameters of the group of symmetry also become 
noncommutative. Thus there are no usual group of transformations, but instead 
there exists only $E_q(2)$-group {\it coaction} on the algebra of 
$q$-coordinates. For clear physical interpretation of quantum symmetries, 
it is desirable to derive transformations of states of a physical system under a coaction 
of a quantum group. Recall, that in the case of Lie algebra-like 
spaces the coordinates form a tensor operator $\Hx_i\ra
\Hx'_i=M_{ij}\Hx_j+b_i=\hU_g\Hx_i\hU_g^{-1}$ and states of the field system
are transformed by the operator $\hU_g$ (we considered the examples of such
transformations for the case of noncommutative Euclidean and Minkowski
planes in section~\ref{stqp}). In the $q$-deformed case, these transformations 
are defined by the generalized Clebsch-Gordan
coefficients, describing decomposition of tensor products of representations
of algebras of functions on quantum spaces and representations of the
corresponding quantum group. In other words, the coaction defines (by duality) 
a map $\cS$ which puts in correspondence to a pair of states on a group 
algebra $G_q$ and on a quantum space $\cX_q$ some new state on the 
quantum space:
\be
\cS\bigg(\ket{\PS},\ket{\p}\bigg)=\ket{\p'}\qq 
\ket{\PS}\in\cH_{G_q}\,,\ \ \ket{\p},\ket{\p'}\in\cH_{\cX_q}\ .\e{short.q-sym}
\ee
Matrix elements of the map $\cS$ are the generalized Clebsch-Gordan
coefficients. We shall consider this new type of symmetry transformations 
in full detail in future communication, while in the rest of the present paper 
we consider ultraviolet behaviour of a field theory on the $q$-plane.

\subsection{Distorted plane waves on a noncommutative quantum plane}

It is well-known that the Casimir operator generating the centrum of 
the enveloping algebra  ${\cal U}_q e(2)$ is given by the formula
\be
C\ =\ {\bar T} \er^{2J} T\ .            \e{pp1}
\ee
On a quantum plane it is the Laplace operator $\Delta = \lambda (C)$ (\cf
\r{2.8a}), or more explicitly,
\be
\Delta \ =\ {\bar \partial}_q q^{2{\bar N}-2N} \partial_q 
=-q^2\bar PP\ .\e{pp2}
\ee
Our aim in this subsection is to find eigenfunctions of $\Delta$ which 
further will be used for construction of a field theory on the $q$-plane.

We search for the functions $\Phi_{kk'} (z,{\bar z})$ on the $q$-plane which
are common eigenfunctions of two commuting operators $\Delta$ and 
$\partial_q$:
\be
\Delta \Phi_{kk'} (z,{\bar z})\ =\ kk'\Phi_{kk'} (z,{\bar z})\ ,\e{pp3}
\ee
\be
\partial_q \Phi_{kk'} (z,{\bar z})\ =\ k \Phi_{kk'} (z,{\bar z})\ .\e{pp4}
\ee
The solutions of this eigenvalue problem have the form
\be
\Phi_{kk'} \ =\ \er^{kz}_{q^{-2}} {\tilde e}^{k'{\bar z}}_{q^2} \ ,\
k,k' \in \C\ .           \e{pp5}
\ee
The first factor is the usual $q$-exponential 
\be
\er^{kz}_{q^{-2}} \ =\ \sum_{n=0}^\infty \frac{(kz)^n}{[n;q^{-2}]!} \ ,\e{pp6}
\ee
and it guarantees the validity of \r{pp4}. The explicit form of the 
second factor is obtained by inserting \r{pp5} into \r{pp4}:
\be
{\tilde e}^{k'{\bar z}}_{q^2} \ =\ \sum_{n=0}^\infty 
\frac{(k'{\bar z})^n q^{-n(n+1)}}{[n;q^2 ]!} \ .  \e{pp7}
\ee

We shall show (\cf \r{pp20}) that the eigenfunctions form a basis of the
Hilbert space $\cH_q$ of square integrable (in an appropriate sense, see below)
``functions'' on the $q$-plane $P^{(2)}_q$. The Laplacian $\D$ is the
hermitian operator (see \r{2.10}, \r{2.16}) and, moreover, the expression in
terms of the conjugate operators $\bar P$ and $P$ in \r{pp2} shows that $(-\D)$
is positive semidefinite. Therefore, we must put in \r{pp5} $k'=-\bar k$. 
Thus, the (properly normalized, see below) distorted plane waves
are
\be
\Phi_k (z,{\bar z})\ =\ q^{-1/2} \er^{kz}_{q^{-2}}
{\tilde e}^{-{\bar k}{\bar z}}_{q^2} \ ,\ k \in \C\ .  \e{pp8}
\ee
They are solutions of the following eigenvalue problem:
\bn
\Delta \Phi_k (z,{\bar z})&=&-k{\bar k} \Phi_k (z,{\bar z})
\ ,\nonumber\\[-1mm]
&& \e{pp9}\\[-1mm]
\partial \Phi_k (z,{\bar z})&=& k \Phi_k (z,{\bar z})\ .\nonumber
\en
We stress that they are {\it not} eigenfunctions of
${\bar \partial}_q$.

Let us now restrict to the case $q>1$ (the case $q<1$ will be described later).
We shall interpret the plane waves $\Phi_k (z,{\bar z})$, $k\in\C$, 
as distributions acting on a suitable set of test functions on a quantum
plane $f(z,{\bar z})$ according to the rule:
\be
(\Phi ,f) = \int^q d{\bar z}dz\ \Phi (z,{\bar z}) f(z,{\bar z})\ .\e{pp10}
\ee
The integral introduced here is the $E_q (2)$ invariant integral which
for $q>1$ can be defined in the following way (see, \eg \cite{ChaichianDbk}):
\be
\int^q d{\bar z}dz\ \er^{-{\bar z}z}_{q^2} z^n {\bar z}^m =
\delta_{mn} [n;q^2 ]!\ ,\ q>1\ .         \e{pp11}
\ee
As for the set of test functions we take the space of functions of the 
form
\be
f(z,{\bar z})\ =\ \er^{-{\bar zz}}_{q^2} 
\sum_{m,n=0}^\infty a_{mn} z^n {\bar z}^m \ ,\ q>1\ ,  \e{pp12}
\ee
with rapidly decreasing coefficients for $n,m \to \infty$.

The Fourier transform of any test function \r{pp12} we define by
\be
{\tilde f}(k,{\bar k})\ =\ \int^q d{\bar z}dz\ f(z,{\bar z}) 
\Phi_k (z,{\bar z})\ .                   \e{pp13}
\ee
In particular, the Fourier transform of a $q$-Gaussian packet can be 
calculated easily, and we recover a $q$-analog of the well-known formula
\be
\alpha^{-2} \er^{-k{\bar k}q^{-2} \alpha^{-2}}_{q^2} \ =\
\int^q d{\bar z}dz\ \er^{-\alpha^2 {\bar z}z}_{q^2} \Phi_k (z,{\bar z})
\ .                            \e{pp14}
\ee

So far we considered all $k\in\C$. On the other hand, using the set of the
operators $\bar P,\ P,\ q^{2(\bar N-N)},\ q^{2(\bar N+N+1)}$, one can easily
find (similar to \r{2.38},\r{2.39}) that the operator $\D=-q^2\bar PP$ has a
discrete spectrum. Thus, the eigenfunctions $\F_k,\ k\in\C$ form an
overcomplete set. To use only complete set of eigenfunctions, we have to
reduce consideration to an appropriate discrete subset, \eg $\F_{q^{2n}},\
n=0,\pm1,\pm2,...\,$. Correspondingly, an integration over variables $k,\bar
k$ should be defined in a sense of Jackson-like integral, with the values of
basic integrals for $q>1$ being as follows (see \cite{ChaichianDbk}, 
sect.~2.2): 
\be
\int_q d{\bar k}dk\ \er^{-k{\bar k}q^{-2}}_{q^2} k^n {\bar k}^m \ =\
q \delta_{mn} [n;q^{-2}]!\ ,\ q>1\ .         \e{pp15}
\ee

The inverse Fourier transform to \r{pp13} is given as 
\be
f(z,{\bar z}) = \int_q d{\bar k}dk\ {\tilde f}(k,{\bar k})
\Phi^\da_k (z,{\bar z}) \ .              \e{pp16}
\ee
In particular, the validity of \r{pp16} can be shown easily for the
$q$-Gaussian packet:
\be
\er^{-\alpha^2 {\bar z}z}_{q^2} \ =\ \int_q d{\bar k}dk\ \alpha^{-2}
\er_{q^2}^{-k{\bar k}q^{-2} \alpha^{-2}} \Phi^\da_k (z,{\bar z}) \ .\e{pp17}
\ee
Analogously, the validity of \r{pp16} can shown straightforwardly for any
test function of the form $\er^{-{\bar z}z}_{q^2} z^n {\bar z}^m$.
This proves that the system of distorted plane waves 
$\Phi_k (z,{\bar z})$, $k=q^{2n},\ n$-integers, is a complete orthogonal set
of functions on the quantum plane normalized to the
$\delta$-function:
\be
\int^q d{\bar z}dz\ \F^\da_k (z,{\bar z})
\Phi_{k'} (z,{\bar z})\ =\ \delta^{(2)}_q (k,k') \ .\e{pp18}
\ee
The symbol on the right hand side is defined as an analog of
$\delta$-function on the quantum plane by the relation
\be
\int_q d{\bar k}'dk'\ {\tilde f}(k',{\bar k}') 
\delta^{(2)}_q (k,k')\ =\ {\tilde f}(k,{\bar k})\ ,  \e{pp19}
\ee
which should be valid for all Fourier transforms 
${\tilde f}(k,{\bar k})$ of any test function.

By ${\cal H}_q$ we denote the Hilbert space of  functions on a scalar
plane for which there exists the scalar product
\be
(f_1 ,f_2 ) \ = \ \int^q d{\bar z}dz\ f^\da_1 (z,{\bar z})
f_2 (z,{\bar z})\ .                  \e{pp20}
\ee
It can be shown straightforwardly that the Laplacian $\Delta$ is
symmetric on a subspace of ${\cal H}_q$ formed by test functions 
\r{pp12}. Then by the construction presented above, the operator 
$\Delta$ can be extended to a self-adjoint operator, with plane 
waves $\Phi_k (z,{\bar z})$, $k\in\C$.

Let us now discuss briefly the case $q<1$. In this case we do not
take the test functions in the form \r{pp12} but as the series
\be
f(z,{\bar z})\ =\ \er^{-z{\bar z}}_{q^{-2}} \sum_{m,n=0}^\infty
a_{nm} {\bar z}^m z^n \ ,\ q<1 \ ,         \e{pp21}
\ee
with rapidly decreasing coefficients. Formulas \r{pp13} and \r{pp16} 
defining the direct and inverse Fourier transforms remain valid 
but we have to properly modified the definition of integrals. Namely,
eqs. \r{pp11} and \r{pp15} should be replaced respectively by
\be
\int^q d{\bar z}dz\ \er^{-z{\bar z}}_{q^{-2}} {\bar z}^m z^n \ =\
\delta_{mn} [m;q^2 ]! \ ,\qq q<1 \ ,        \e{pp22}
\ee
and
\be
\int_q d{\bar k}dk\ \er^{-k{\bar k}q^2}_{q^{-2}} k^m {\bar k}^n 
\ =\ q^{-1} \delta_{mn} [n;q^2 ]!\ ,\qq q<1 \ .  \e{pp23}
\ee
We do not introduce new symbols for the integrals \r{pp22} and \r{pp23},
since they correspond to the same Jackson integrals as \r{pp11} and
\r{pp15}, respectively. This can be seen from the matching formulas
between both pairs of integrals. The matching formula between \r{pp11}
and \r{pp22} is:
\be
\int^q d{\bar z}dz\ \er^{-\beta^2 {\bar z}z}_{q^2} 
\er^{-\alpha^2 z{\bar z}}_{q^{-2}}\ =\ \frac{1}{\alpha^2 +\beta^2}\ 
=\ \int^q d{\bar z}dz\ \er^{-\alpha^2 z{\bar z}}_{q^{-2}}
\er^{-\beta^2 {\bar z}z}_{q^2} \ .  \e{pp24}
\ee
We have expanded in both sides the second exponent and 
used the definitions \r{pp11} and \r{pp12}. The result on the left hand side
is justified for $\alpha < \beta$, whereas on the right hand side for
$\alpha > \beta$. A similar matching formula is valid between
integrals \r{pp15} and \r{pp23}:
\be
\int_q d{\bar k}dk\ \er^{-\beta^2 k{\bar k} q^{-2}}_{q^2}
\er^{-\alpha^2 k{\bar k}q^2}_{q^{-2}}\ =\ \frac{1}{q\alpha^2 +
q^{-1}\beta^2}\ =\ \int_q d{\bar k}dk\ \er^{-\alpha^2 k
{\bar k}q^2}_{q^{-2}} \er^{-\beta^2 k{\bar k}q^{-2}}_{q^2} \ .\e{pp25}
\ee
Both sides have been evaluated in the similar way as before,
however now we have used \r{pp15} and \r{pp23}. Again, left hand side is
valid for $\alpha < \beta$ and the right hand side for $\beta >\alpha$.

Notice, that the whole analysis presented above can be repeated
with the second system of distorted plane waves given as
\be
\Psi_k (z,{\bar z})\ =\ q^{-1/2} \er^{-{\bar k}{\bar z}}_{q^2}
{\tilde e}^{q^{-2}kz}_{q^{-2}} \ ,\ k \in\C\ ,  \e{pp26}
\ee
which solve the eigenvalue problem:
\[
\Delta \Psi_k (z,{\bar z})\ =\ -k{\bar k}\Psi_k (z,{\bar z})\ ,
\]
\be
{\bar \partial}_q \Psi_k (z,{\bar z})\ =\ k\Psi_k (z,{\bar z})\ .\e{pp27}
\ee
There is also an alternative possibility to use $q$-deformed
spherical waves on a quantum plane which are eigenfunctions of
the Laplacian $\Delta$ and the angular momentum operator $J$.

\subsection{The field action and the tadpole contribution}

We define on the quantum plane the field action for a complex  
field $\vf (z,{\bar z})$ in a standard way
\be
S[\vf ,\vf^\da] = \int^q d{\bar z}dz\ [\vf^\da (z,{\bar z})
(-\Delta +m^2 )\vf (z,{\bar z})\ -\ V(\vf ,\vf^\da )]\ .  \e{pp28}
\ee
We expand the field $\vf (z,{\bar z})$ into the plane waves \r{pp8}
\be
\vf (z,{\bar z})\ =\ \int_q d{\bar k}dk\, a(k,{\bar k})
\Phi_k (z,{\bar z})\ .               \e{pp29}
\ee
Inserting this into \r{pp28} we can diagonalize the free field 
action:
\bn
S_0 [\vf ,\vf^\da]& =& \int^q d{\bar z}dz\ \vf^\da (z,{\bar z})
(-\Delta +m^2 )\vf (z,{\bar z})\nbr  
&=& \int_q d{\bar k}dk\ \bar a (k,{\bar k})
(k{\bar k}+m^2 ) a(k,{\bar k})\ .         \e{pp30}
\en
The last formula is valid due to the completeness and 
orthogonality properties of the distorted plane waves.

{}From \r{pp30} we see that the free field propagator
in the $k$-representation is given as
\be
{\tilde G}(k,{\bar k})\ =\ \frac{1}{k{\bar k}+m^2}\ . \e{pp31}
\ee

Since, the propagator ${\tilde G}(k,{\bar k})$ behaves like
$(k{\bar k})^{-1}$ for large $k{\bar k}$, the UV-divergencies
has to be expected. Indeed, let us consider the scalar field on the $q$-plane
in a constant external field $F$, so that the interaction term has the simple
form
$$
V(\vf ,\vf^\da )=F\vf^\da(z,{\bar z})\vf(z,{\bar z})\ ,\qq F=\const\ .
$$
Then the tadpole contribution is proportional to
the divergent Jackson integral
\be
\Gamma = \int_q d{\bar k}dk \frac{1}{k{\bar k}+m^2}=\infty\ .    \e{pp32}
\ee

In order to prove \r{pp32} we shall perform the evaluation of 
the tadpole contribution with the regularized propagator 
\be
{\tilde G}^{reg}_s \ = \ m^{2s} \int^q d{\bar z}dz\ 
\er^{-m^2{\bar z}z}_{q^2} \er^{-k{\bar k}z{\bar z}}_{q^{-2}}
(z{\bar z})^s \ .                         \e{pp33}
\ee
For $s>-1$, this is a well-defined Jackson integral behaving
asymptotically as $(k{\bar k})^{-1-s}$. For $s=0$ it reduces
to \r{pp31} as can be seen from \r{pp24}. Inserting \r{pp33} into the
tadpole integral and interchanging the order of integrations,
the regularized tadpole contribution can be calculated 
straightforwardly,
\bn
\Gamma^{reg}_s & =& m^{2s} \int^q d{\bar z}dz  
\er^{-m^2{\bar z}z}_{q^2} (z{\bar z})^s \int_q d{\bar k}dk
\er^{-k{\bar k}z{\bar z}}_{q^{-2}} \nbr
&=& \int^q d{\bar z}dz \er^{-{\bar z}z}_{q^2} 
(z{\bar z})^{s-1} \ =\ \frac{1}{2} \Gamma_{q^2} (s)\ .   \e{pp34}
\en 
Here, $\Gamma_{q^2} (s)$ is a $q$-deformed $\Gamma$-function
which is singular at $s=0$, similarly to the usual 
$\Gamma$-function (repeating the indicated calculations
in the standard case we recover \r{pp34} with the ordinary
$\Gamma$-function).

In the case of the external field $F(z,\bar z)=\sum_{n,m=0}^\infty
a_{mn}\bz^nz^m$ depending on coordinates, for commutative plane there appears a 
divergent tadpole contribution of the form $\wt F(0,0)\G_{com}$, where 
$\wt F(k,\bar k)$ is the \fr transform of the external field and $\G_{com}$ 
is the tadpole integral on the commutative plane. 
On the noncommutative plane we have the interaction action as follows
\bnn
S_{int}&=&\int^q d{\bar z}dz\,\vf^\da(z,{\bar z})
F({\bar z},z)\vf(z,{\bar z})\nbr
&=&\int_q d{\bar k}dk\,g(k_1,\bar k_1,k_2,\bar k_2)
\wt\vf^\da(k_1,{\bar k_2})\wt\vf(k_2,{\bar k_2})\ .
\enn
Taking into account the explicit form \r{pp8} of plane waves and their
orthogonality \r{pp18}, the vertex $g(k_1,\bar k_1,k_2,\bar k_2)$ 
can be cast into the form
\bnn
g(k_1,\bar k_1,k_2,\bar k_2)&=&\int^qd{\bar z}dz\,
\Phi^\da_{k_1} (z,{\bar z})F(z,\bz)\Phi_{k_2} (z,{\bar z})\nbr
&=&F(\pa_{k_1},\pa_{\bar k_2})\delta^{(2)}_q (k_1,k_2)\ 
\enn
(where the derivatives are given by $\pa_{k}k^n\equiv[n;q^{-2}]k^{n-1}$
and $\pa_{\bar k}\bar k^n\equiv[n;q^{-2}]\bar k^{n-1}$). The tadpole
contribution is given as
$$
\int_q d{\bar k}dk\,\frac{g(k,\bar k,k,\bar k)}{k\bar k+m^2}\ .
$$
In the commutative case the numerator $g(k,\bar k,k,\bar k)=\wt F(0,0)$ is
constant leading to the divergent tadpole contribution. In the noncommutative
case we obtain even more divergent contributions which cannot cancel each other
for a general field $F(z,\bz)$ (due to arbitrariness of the expansion 
coefficients $a_{mn}$). 
They are proportional to $\log q$ and disappear in the commutative limit. 
 
We conclude that the $q$-deformation of a plane does
not lead to the desired UV-regularization. 

\section{Conclusion}

We have shown that transition to a noncommutative space-time does not 
necessarily lead to an ultraviolet regularization of the quantum field theory
constructed in this space, at least in the most natural way of introducing
noncommutativity as we have performed in this paper. 
In particular, QFT on noncommutative planes with the
Heisenberg-like commutation relations for coordinates and the deformed plane
with the quantum $E_q(2)$-symmetry still contain divergent tadpoles. However,
in general, theories which have the same UV-behaviour on  classical spaces may
acquire essentially different properties after the quantization. The reason is
that quantization procedure is highly sensitive to the topology of the manifold
under consideration. Thus, while in the case of  classical space-time the
theories on a sphere, cylinder or plane have UV-divergences, in the case of
noncommutative space-time the two-dimensional theories on the fuzzy sphere
\cite{Hoppe,GKP3} and on the quantum cylinder do not have divergences at all.
This can be traced to the compactness properties of the space-time in question:
\biz
\item[-]
In the case of a fuzzy sphere, models contain a finite number of modes and thus
all the usual integrations are replaced by final sums
and, consequently, no UV-divergences can appear.
\item[-]
In the case of a cylinder, a priori one can not claim whether the 
quantum field theory is finite. However, the non-commutativity of the space-time together with the compactness of the space
(circle) lead to the intrinsic cut-off in the energy modes. This guarantees the
removal of UV-divergences in the two-dimensional case.
\item[-]
On a noncommutative plane (whose commutative limit is noncompact in both
directions) with Heisenberg-like or even with deformed commutation relations, 
the noncommutativity of the space-time does not lead to an UV-regular theory.
\eiz

We conclude with the following general picture:
\biz
\item[-]
the noncommutativity itself does not guarantee the removal of UV-divergences;
\item[-]
global topological restrictions are needed - namely,  at most one
dimension (time) is allowed to be noncompact, in order to achieve the removal
of UV-divergences of a quantum field theory formulated in a noncommutative
space-time of arbitrary dimensions.
\eiz

It is of great interest to investigate the problem further in order to find out
whether there exists any other, than the one presented in this paper, way of
introducing noncommutativity so that it would remove the UV-divergences even in
the case of fully noncompact space-times.

\vspace{5mm}

{\bf Acknowledgements} 

The financial support of the Academy of Finland under the Projects No. 37599
and 44129 is greatly acknowledged.
A.D.'s work was partially supported also by RFBR-98-02-16769 grant and P.P.'s
work by VEGA project 1/4305/97.

\end{document}